\documentclass[twocolumn]{aastex631} % twocolumn, linenumbers, manuscript
\usepackage{amsmath}
\renewcommand{\arraystretch}{1.5}

\newcommand{\magphys}{\hbox{\sc magphys}}
\newcommand{\av}{\hbox{$A_{\rm V}$}}
\newcommand{\fobs}{\hbox{$f_{\rm obs}$}}
\newcommand{\sigmstar}{\hbox{$\Sigma_{M_*}$}}
\newcommand{\sigsfr}{\hbox{$\Sigma_{\rm SFR}$}}
\newcommand{\sigcii}{\hbox{$\Sigma_{L_{\rm [CII]}}$}}
\newcommand{\ldust}{\hbox{$L_{\rm dust}$}}

\shorttitle{ALMA$-$CRISTAL: resolved $4<z<6$ star-forming galaxies}
\shortauthors{Li et al.}

\begin{document}

\title{The ALMA$-$CRISTAL survey:
Spatially-resolved star formation activity and dust content in $4<z<6$ star-forming galaxies}
\author[0000-0002-8184-5229]{Juno Li} %%% juno.li.research@gmail.com
\affiliation{International Centre for Radio Astronomy Research (ICRAR), \\The University of Western Australia, M468, 35 Stirling Highway, Crawley, WA 6009, Australia}
\affiliation{International Space Centre (ISC), The University of Western Australia, M468, 35 Stirling Highway, Crawley, WA 6009, Australia}
\affiliation{ARC Center of Excellence for All Sky Astrophysics in 3 Dimensions (ASTRO 3D), Australia}
\author[0000-0001-9759-4797]{Elisabete Da Cunha}
\affiliation{International Centre for Radio Astronomy Research (ICRAR), \\The University of Western Australia, M468, 35 Stirling Highway, Crawley, WA 6009, Australia}
\affiliation{ARC Center of Excellence for All Sky Astrophysics in 3 Dimensions (ASTRO 3D), Australia}
%%%% Builders %%%%%
\author[0000-0003-3926-1411]{Jorge González-López}
\affiliation{Las Campanas Observatory, Carnegie Institution of Washington, Casilla 601, La Serena, Chile}
\affiliation{Instituto de Estudios Astrofísicos, Facultad de Ingeniería y Ciencias, Universidad Diego Portales, Av. Ejército 441, Santiago, Chile}
\author[0000-0002-6290-3198]{Manuel Aravena}
\affiliation{Instituto de Estudios Astrofísicos, Facultad de Ingeniería y Ciencias, Universidad Diego Portales, Av. Ejército 441, Santiago, Chile}
\author[0000-0001-9419-6355]{Ilse De Looze}
\affiliation{Sterrenkundig Observatorium, Ghent University, Krijgslaan 281 S9, B-9000 Ghent, Belgium}
\author[0000-0003-4264-3381]{N. M. Förster Schreiber}
\affiliation{Max-Planck-Institut für Extraterrestische Physik (MPE), Giessenbachstr., 85748, Garching, Germany}
\author[0000-0002-2775-0595]{Rodrigo Herrera-Camus}
\affiliation{Departamento de Astronom\'ia, Universidad de Concepci\'on, Barrio Universitario, Concepci\'on, Chile}
\affiliation{Department of Physics \& Astronomy, University College London, Gower Street, London WC1E 6BT, UK}
\author[0000-0003-3256-5615]{Justin Spilker}
\affiliation{Department of Physics and Astronomy and George P. and Cynthia Woods Mitchell Institute for Fundamental Physics and Astronomy, Texas A\&M University, College Station, TX, USA}
\author[0000-0001-9728-8909]{Ken-ichi Tadaki}
\affiliation{Faculty of Engineering, Hokkai-Gakuen University, Toyohira-ku, Sapporo 062-8605, Japan}
% \author{Roberto J. Assef}
% \affiliation{Instituto de Estudios Astrof\'isicos, Facultad de Ingenier\'ia y Ciencias, Universidad Diego Portales, Av.  Ej\'ercito Libertador 441, Santiago, Chile}
% \author{Rychard Bouwens}
% \affiliation{Leiden Observatory, Leiden University, NL-2300 RA Leiden, Netherlands}
\author[0000-0003-0057-8892]{Loreto Barcos-Munoz}
\affiliation{National Radio Astronomy Observatory, 520 Edgemont Road, Charlottesville, VA 22903, USA}
\affiliation{Department of Astronomy, University of Virginia, 530 McCormick Road, Charlottesville, VA 22903, USA}
 
%%%%%%%%Co-authors%%%%%%%%
\author[0000-0003-4569-2285]{Andrew J. Battisti}
\affiliation{International Centre for Radio Astronomy Research (ICRAR), \\The University of Western Australia, M468, 35 Stirling Highway, Crawley, WA 6009, Australia}
\affiliation{ARC Center of Excellence for All Sky Astrophysics in 3 Dimensions (ASTRO 3D), Australia}
\affiliation{Research School of Astronomy and Astrophysics, Australian National University, Cotter Road, Weston Creek, ACT 2611, Australia}
\author[0000-0002-3272-7568]{Jack E. Birkin}
\affiliation{Department of Physics and Astronomy and George P. and Cynthia Woods Mitchell Institute for Fundamental Physics and Astronomy, Texas A\&M University, College Station, TX, USA}
\author[0000-0003-3917-1678]{Rebecca A. A. Bowler}
\affiliation{Jodrell Bank Centre for Astrophysics, Department of Physics and Astronomy, School of Natural Sciences, The University of Manchester, Manchester, M13 9PL, UK}
\author[0000-0002-3324-4824]{Rebecca Davies}
\affiliation{Centre for Astrophysics and Supercomputing, Swinburne Univ. of Technology, PO Box 218, Hawthorn, VIC, 3122, Australia}
\author[0000-0003-0699-6083]{Tanio Díaz-Santos}
\affiliation{Institute of Astrophysics, Foundation for Research and Technology-Hellas, GR-71110, Heraklion, Greece}
\affiliation{School of Sciences, European University Cyprus, Diogenes Street, Engomi, 1516 Nicosia, Cyprus}
\author[0000-0002-9400-7312]{Andrea Ferrara}
\affiliation{Scuola Normale Superiore, Piazza dei Cavalieri 7, 56126 Pisa, Italy}
\author[0000-0003-0645-5260]{Deanne B. Fisher}\affiliation{Centre for Astrophysics and Supercomputing, Swinburne Univ. of Technology, PO Box 218, Hawthorn, VIC, 3122, Australia}
\affiliation{Leiden Observatory, Leiden University, NL-2300 RA Leiden, the Netherlands}
\affiliation{Max-Planck Institute for Astrophysics, Karl Schwarzschildstrasse 1, 85748, Garching, Germany}
\author[0000-0001-6586-8845]{Jacqueline Hodge}
\affiliation{Leiden Observatory, Leiden University, NL-2300 RA Leiden, the Netherlands}
\author[0000-0002-2634-9169]{Ryota Ikeda}
\affiliation{Department of Astronomy, School of Science, SOKENDAI (The Graduate University for Advanced Studies), 2-21-1 Osawa, Mitaka, Tokyo 181-8588, Japan}
\affiliation{National Astronomical Observatory of Japan, 2-21-1 Osawa, Mitaka, Tokyo 181-8588, Japan}
\author[0000-0001-5289-3291]{Meghana Killi}
\affiliation{Instituto de Estudios Astrofísicos, Facultad de Ingeniería y Ciencias, Universidad Diego Portales, Av. Ejército 441, Santiago, Chile}
\author[0000-0001-7457-4371]{Lilian Lee}
\affiliation{Max-Planck-Institut für Extraterrestische Physik (MPE), Giessenbachstr., 85748, Garching, Germany}
\author[0000-0001-9773-7479]{Daizhong Liu}\affiliation{Max-Planck-Institut für Extraterrestische Physik (MPE), Giessenbachstr., 85748, Garching, Germany}\affiliation{Purple Mountain Observatory, Chinese Academy of Sciences, 10 Yuanhua Road, Nanjing 210023, China}
\author[0000-0003-0291-9582]{Dieter Lutz}
\affiliation{Max-Planck-Institut für Extraterrestische Physik (MPE), Giessenbachstr., 85748, Garching, Germany}
\author[0000-0001-7300-9450]{Ikki Mitsuhashi}
\affiliation{National Astronomical Observatory of Japan, 2-21-1 Osawa, Mitaka, Tokyo 181-8588, Japan}
\author[0000-0002-7314-2558]{Thorsten Naab}
\affiliation{Max-Planck Institute for Astrophysics, Karl Schwarzschildstrasse 1, 85748, Garching, Germany}
\author{Ana Posses}
\affiliation{Instituto de Estudios Astrofísicos, Facultad de Ingeniería y Ciencias, Universidad Diego Portales, Av. Ejército 441, Santiago, Chile}
\author[0000-0003-1682-1148]{Monica Relaño}
\affiliation{Dept. Física Teórica y del Cosmos, Universidad de Granada, Spain}
\affiliation{Instituto Universitario Carlos I de Física Teórica y Computacional, Universidad de Granada, 18071, Granada, Spain}
\author[0000-0001-6629-0379]{Manuel Solimano}
\affiliation{Instituto de Estudios Astrofísicos, Facultad de Ingeniería y Ciencias, Universidad Diego Portales, Av. Ejército 441, Santiago, Chile}
\author[0000-0003-4891-0794]{Hannah Übler}
\affiliation{Kavli Institute for Cosmology, University of Cambridge, Madingley Road, Cambridge CB3 0HA, UK}
\affiliation{Cavendish Laboratory, University of Cambridge, 19 JJ Thomson Avenue, Cambridge, CB3 0HA, UK}
\author[0009-0001-4144-9635]{Stefan Anthony van der Giessen}
\affiliation{Sterrenkundig Observatorium, Ghent University, Krijgslaan 281-S9, 9000 Ghent, Belgium}
\affiliation{Dept. Fisica Teorica y del Cosmos, Universidad de Granada, Spain}
\author[0000-0002-5877-379X]{Vicente Villanueva}
\affiliation{Departamento de Astronom\'ia, Universidad de Concepci\'on, Barrio Universitario, Concepci\'on, Chile}

\begin{abstract}
\textcolor{black}{Using a combination of HST, JWST, and ALMA data, we perform spatially resolved spectral energy distributions (SED) fitting of fourteen $4<z<6$ UV-selected main-sequence galaxies targeted by the [CII] Resolved ISM in Star-forming Galaxies with ALMA (CRISTAL) Large Program.} We consistently model the emission from stars and dust in $\sim$0.5-1kpc spatial bins to obtain maps of their physical properties. We find no offsets between the stellar masses (M$_*$) and star formation rates (SFRs) derived from their global emission and those from adding up the values in our spatial bins, suggesting there is no bias of outshining by young stars on the derived global properties. \textcolor{black}{We show that ALMA observations are important to derive robust parameter maps because they reduce the uncertainties in \ldust\ (hence Av and SFR).} Using these maps we explore the resolved star-forming main sequence for z$\sim$5 galaxies, finding that this relation persists in typical star-forming galaxies in the early Universe. We find less obscured star formation where the M$_*$ (and SFR) surface densities are highest, typically in the central regions, contrary to the global relation between these parameters. We speculate this could be caused by feedback driving gas and dust out of these regions. However, more observations of infrared luminosities with ALMA are needed to verify this. Finally, we test empirical SFR prescriptions based on the UV+IR and [CII] line luminosity, finding they work well at the scales probed ($\sim$kpc). Our work demonstrates the usefulness of joint HST, JWST, and ALMA resolved SED modeling analyses at high redshift.
\end{abstract}

% \keywords{galaxies: high-redshift -- galaxies: evolution -- galaxies: stellar content -- galaxies: ISM -- galaxies: star formation}

\section{Introduction}\label{sec:intro}
The cosmic epoch that followed the reionization of the Universe, about 1~Gyr after the Big Bang ($z\sim4-6$), is critical to our understanding of structure formation and galaxy evolution. The Universe had just become transparent to ultraviolet (UV) photons, and the cosmic star formation history was rapidly ramping up (e.g., \citealt{Madau2014}). Surveys with the Hubble Space Telescope (HST) detected large samples of star-forming galaxies in this active period of cosmic history (e.g., \citealt{Bouwens2007,Bouwens2015}), giving us insight into the evolution of the UV luminosity function and star formation in the Universe at those epochs.

Observations of high-redshift galaxies with the HST show disturbed morphologies and clumpy structures (e.g., \citealt{Elmegreen2007,Forster2011}). However, our view of high-redshift galaxies with the HST is incomplete, as it only probes the rest-frame UV emission, dominated by recent ($\lesssim100$ Myr) star formation, and is strongly affected by dust attenuation.
Thanks to the advent of the James Webb Space Telescope (JWST) and the Atacama Large Millimetre/submillimeter Array (ALMA), we can now access the rest-frame optical/near-infrared emission of early galaxies, probing the bulk of their stellar mass, as well as the dust emission tracing obscured star formation, and tracers of the gas in their interstellar medium. Synergies between these powerful telescopes offer exciting prospects to deepen our understanding of the stellar populations and multi-phase interstellar medium in high-redshift galaxies. Observations in the (rest-frame) infrared are particularly important,  since ALMA studies have shown that dust-obscured star formation at high redshifts, even in the epoch of reionization, may be more prevalent than was previously thought (e.g.,~\citealt{Inami2022,Algera2023,Barrufet2023,Mitsuhashi2023,Bowler2024}).

Physically-motivated spectral energy distribution (SED) models are critical in the interpretation of multi-wavelength observations of galaxies. 
Various SED fitting models have been developed (see, e.g., \citealt{Walcher2011,Conroy2013,Pacifici2023} for reviews) that are now routinely used to interpret the integrated ultraviolet-to-infrared emission of galaxies in large surveys, both in the nearby and the distant Universe (e.g., \citealt{Smith2012,magphys2015,Salim2016,Driver2018,Faisst2020,Dudzeviciute2020,Thorne2021,ceers2023}). Recent observations with the JWST, with its unprecedented resolution and sensitivity at wavelengths longer than HST, have encouraged spatially-resolved SED fitting of $z>3$ galaxies (e.g., \citealt{Abdurrouf2023,GimenezArteaga2023,Song2023,Smail2023}; though see also, e.g., \citealt{Wuyts2011,Wuyts2012,Wuyts2013,Lang2014} for pre-JWST resolved studies at $z\sim1-2$). This allows us to probe the spatial distribution of physical parameters in galaxies, a great improvement as high resolution simulations have illustrated that different physical components in galaxies may not coincide spatially, and their spatial distribution may be different to the light distribution in any particular wavelength \citep{Cibinel2015, cochrane2019,Popping2022}. Such spatially-resolved SED studies can also enable more direct comparisons between observations and recent numerical simulations that make detailed predictions on the spatial distribution of physical properties in high-redshift galaxies (e.g., \citealt{Pallottini2022}).%, allowing us to, for example, test feedback prescriptions in the simulations.

Some spatially resolved studies have found that integrated SED fitting may underestimate the stellar mass in galaxies with high specific star formation rates (sSFR) due to young stars outshinning older stellar populations \citep{Wuyts2012,Sorba2015,Sorba2018}.
%maybe leave this out and include later in discussion
%The spatial resolution of the resolved studies also influence the results. \citet{Sorba2015} found that the offset in stellar mass diminish when the physical scale spatial resolution increase to $\sim$3 kpc. \citet{Wuyts2012} also found that there is no significant difference when they used the Voronoi two-dimensional binning technique to ensure that the minimum S/N of each spatial element was 10.
On the other hand, in a recent JWST study, \cite{Song2023} were able to reproduce the inconsistencies between the resolved and integrated SED fitting approaches by deliberately dropping filters that probe rest-frame wavelengths longer than $1\,\mu$m \textcolor{black}{ which capture the old stellar population which contribute bulk of the stellar mass}. In other words, they find no offset in stellar mass when including filters with rest-frame wavelength longer than $1\,\mu$m in the SED fitting, demonstrating the importance of including rest-frame near-infrared observations enabled by JWST. However, spatially-resolved SED fitting including the dust emission in the mid-/far-infrared has been less explored, due to the lack of matched-resolution dust continuum observations for most samples (though see \citealt{Abdurrouf2022} for an example at low redshift). This may put into question the recovery of the intrinsic physical parameters in the galaxies (such as the stellar mass and star formation rate), since without infrared emission, dust attenuation corrections become more uncertain.

High resolution observations of the dust continuum at high-redshifts ($z>2$) are possible with ALMA (and also NOEMA, e.g. \citealt{Hodge2015,Hodge2016,Hodge2019}), which can match and even surpass the resolution of JWST, and are now becoming available for some galaxy samples. One of those samples is targeted by the [CII] Resolved ISM in Star-forming Galaxies with ALMA (CRISTAL) Large Program (2021.1.00280.L; PI. R. Herrera-Camus), which includes 19 UV-detected, main sequence star-forming galaxies at $4<z<6$.
%These results demonstrated that there are still plenty of rooms to explore on how to perform spatially resolved SED fitting and what new understanding of galaxy it would present.
In this paper, we perform $\sim$kpc-resolution SED fitting of CRISTAL galaxies observed with HST, JWST and ALMA. We develop a method to match the observations in angular resolution and prepare them to input into the \magphys\ SED modeling code \citep{magphys2008,magphys2015}. We obtain physical parameter maps for our sources, which we analyse in detail, and we compare our results with those from unresolved SED fitting. We highlight the importance of having far-infrared observations of the dust continuum in measuring spatially-resolved physical parameters by comparing the results with and without ALMA observations. Finally, we use our physical parameter maps of the CRISTAL galaxies to explore well-known global correlations on sub-kpc scales for the first sample of $z\sim5$ unlensed, main-sequence galaxies.

This paper is structured as follows. We describe our sample selection and data reduction in Section~\ref{sec:data}, and our method to match multi-wavelength observations and fit the SEDs in Section~\ref{sec:method}. We present our resolved parameter maps, explore the resolved star-forming main sequence, and the relation between stellar mass and fraction of obscured star formation rate in Section~\ref{sec:results}. In Section~\ref{sec:discussion}, we discuss how the resolved method differs from unresolved, the impact of having ALMA observations, the radial trends in dust obscuration in our sources, and the usefulness of empirical star formation rate indicators on resolved scales. Finally, we summarize our main findings in Section~\ref{sec:conclusion}.
Throughout this work, we assume a flat universe with cosmological parameters $\Omega_{\rm M}=0.3$, $\Omega_{\Lambda}=0.7$, and $\rm H_0=70\,km\,s^{-1}\,Mpc^{-1}$ (i.e., 1\,\arcsec\ corresponds to 6.28\,kpc at $z=5$), and we adopt a \cite{Chabrier2003} initial mass function (IMF).

\section{Observations and data reduction}\label{sec:data}
\subsection{Sample}\label{sec:sample}

The CRISTAL Large Program targets galaxies selected from a primary sample of 118 [CII]-detected galaxies at $4<z<6$ in the ALMA Large Program to Investigate C$^+$ at Early Times (ALPINE survey; 2017.1.00428.L; PI: O. Le F\`evre; \citealt{LeFevre2020}). CRISTAL followed-up 19 UV-detected by HST, star-forming main sequence galaxies from ALPINE in order to obtain higher-resolution images of their [CII] emission and dust continuum with ALMA. All CRISTAL targets are detected in multiple HST images at rest-frame UV and/or optical.
The selection criteria are based on physical parameters inferred from integrated SED fitting \citep{Faisst2020,Bethermin2020}, including having specific star formation rates (sSFR) within a factor of 3 from the star-forming main sequence at $z\sim5$, and having stellar masses $\rm M_* > 10^{9.5}\rm M_{\odot}$.

CRISTAL includes six additional galaxies (CRISTAL+) in the COSMOS field (HZ4, HZ7, HZ10, DC818760, DC873756, VC8326) that also have ALMA data with similar spatial resolutions and sensitivities as the CRISTAL Large Program (2018.1.01359.S and 2019.1.01075.S, PI: Manuel Aravena; 2018.1.01605.S, PI: Rodrigo Herrera-Camus; 2019.1.00226.S, PI: Edo Ibar). The full CRISTAL survey consists of a total of 25 fields, for which a total of 36 galaxies were identified by either resolving the main target into multiple objects or serendipitously detecting other sources. Details of the survey design and observational setup are presented in Herrera-Camus et al. (in prep.).

Among 36 galaxies identified in the CRISTAL survey, 19 are detected in 
the rest-frame 158$\mu$m dust continuum \citep{Mitsuhashi2023}.
Here we study 14 of those 19 galaxies which were observed by JWST with NIRCam as of January 2024. We adopt the classification of dust detection and naming convention in \cite{Mitsuhashi2023}.
% Among the 25 CRISTAL galaxy fields, 19 were observed by JWST with NIRCam as of January 2024, and 17 galaxies are resolved and detected both in [CII] and in the rest-frame 158$\mu$m dust continuum, with peak dust emission SNR$>3$.
% We selected 14 out of these 16 galaxies as our sample in this work, excluding C14 which the SNR$>3$ peaks are smaller than the beam size showing a hint that the FIR flux may be resolved out.
% The 14 fields have comparable $\sim$kpc angular resolution observations from rest-frame UV (HST), optical (JWST/NIRCam) to dust continuum at 158$\mu$m.
Details about the 14 CRISTAL targets studied in this work can be found in Table~\ref{tab:sample}.

\begin{table*}
\centering
    \begin{tabular}{lllllll}
        \hline
        CRISTAL ID& Alternative name &  $z_{\rm [CII]}$ & R.A. (h:m:s)& Dec (d:m:s) & $\rm log(M_*/M_{\odot})$ & log(SFR/$\rm M_{\odot}yr^{-1}$) \\%& $S_{\rm 850\mu m}/\rm{mJy}$\\
        \hline
        \multicolumn{7}{c}{{\bf CRISTAL main sample}} \\
        CRISTAL-02& DC\_848185 & 5.294 & 10:00:21.50 &+02:35:11.08& $10.30\pm0.28$ & $2.25\pm0.42$ \\
        CRISTAL-03& DC\_536534 & 5.689 & 09:59:53.26& +02:07:05.42 & $10.40\pm0.29$ & $1.79\pm0.31$ \\%& $0.045\pm0.031$ \\
        CRISTAL-04a& VC\_5100822662 & 4.520 & 09:58:57.91 & +02:04:51.48 & $10.15\pm0.29$ & $1.89\pm0.21$ \\% $0.26\pm0.12$ \\
        % CRISTAL-04b & 4.520 & 149.7414 & 2.0813 & $8.91\pm0.55$ & $0.63\pm0.37$ \\%
        CRISTAL-06a& VC\_5100541407 & 4.562 & 10:01:00.93 & +01:48:33.84& $10.09\pm0.30$ & $1.62\pm0.34$ \\%& $0.37\pm0.13$ \\
        % CRISTAL-06b & 4.562 & 150.2542 & 1.8097	& $9.19\pm0.46$ &	$1.07\pm0.33$ \\%& $0.37\pm0.13$ \\
        CRISTAL-07ab& DC\_873321 & 5.154 & 10:00:04.06& +02:37:35.76&	$10.00\pm0.33$ & $1.89\pm0.26$ \\%  $0.17\pm0.067$ \\
        % CRISTAL-08 4.43& 53.0793& -27.8771& $9.85\pm0.36$& $1.88\pm0.23$& \\
        CRISTAL-09& DC\_519281 & 5.575 & 09:59:00.91 & +02:05:27.60& $9.84\pm0.39$ &	$1.51\pm0.32$ \\%  $0.17\pm0.067$ \\
        CRISTAL-11& DC\_630594 & 4.439 & 10:00:32.62 &	+02:15:28.40&	$9.68\pm0.33$ &	$1.57\pm0.31$ \\%  $0.22\pm0.093$ \\
        % CRISTAL-12 & & & & & \\
        CRISTAL-13& VC\_5100994794 & 4.579 & 10:00:41.16 &+02:17:14.13&	$9.65\pm0.34$ &	$1.51\pm0.41$ \\%  $0.058\pm0.058$\\
        % CRISTAL-16 & & & & & \\
        CRISTAL-19& DC\_494763 & 5.233 & 10:00:05.11 & +02:03:12.11 &	$9.51\pm0.36$ &	$1.45\pm0.36$ \\%  $0.084\pm0.033$\\
        \hline
        \multicolumn{7}{c}{{\bf CRISTAL+ sample}} \\
        CRISTAL-21& HZ7 & 5.255 & 09:59:30.48 & +02:08:02.76&	$10.11\pm0.32$ & $1.80\pm0.32$ \\%  $0.22\pm0.086$\\
        CRISTAL-24& DC\_873756 & 4.546 & 10:00:02.71& +02:37:40.20 &	$10.53\pm0.08$ & $0.94\pm0.95$ \\%  $0.22\pm0.086$\\
        CRISTAL-25& VC\_5101218326 & 4.573 & 10:01:12.50 &	+02:18:52.72&	$10.90\pm0.32$ & $2.75\pm0.29$ \\%  $0.22\pm0.086$\\
        \hline
        \multicolumn{7}{c}{{\bf Serendipitously detected sources}} \\
        CRISTAL-01b& -- & 4.530& 10:00:54.77& +02:34:28.20& $9.81\pm0.34$& $1.71\pm0.29$\\
        CRISTAL-07c& -- & 5.155	& 10:00:03.22& +02:37:37.56&	$10.21\pm0.35$ & $1.92\pm0.41$ \\%  $0.17\pm0.067$ \\
        \hline
    \end{tabular}
    \caption{The sample of CRISTAL galaxies used in this work. The stellar masses ($M_\ast$) and star formation rates (SFR) are obtained from \citet{Mitsuhashi2023}, and are based on integrated SED studies (which did not include the ALMA fluxes).
    %The last column shows the flux detected in ALMA band 7 measured in this work.
    }
    \label{tab:sample}
\end{table*}

\subsection{ALMA Band 7 observations}\label{sec:alma}
The CRISTAL observations and ALPINE archival observations were taken in Band 7, targeting the [CII] line ($\nu_\mathrm{rest}=1900.54$~GHz), using a range of ALMA antenna configurations. ALPINE observations were taken using the most compact configurations (C43-1 or C43-2), to prioritize detection over spatial resolution, achieving an average beam size of $0.85\arcsec\times1.13\arcsec$ \citep{LeFevre2020}. CRISTAL observations combine long baselines (C43-5 or C43-6; typical beam sizes $0.1\arcsec-0.3\arcsec$) and short baselines (C43-1 or C43-2; typical beam sizes $0.7\arcsec-1.0\arcsec$) array configurations, such that the observations are sensitive both to small scale structures and extended emission (more details in Herrera-Camus et al., in prep). Visibility data from ALPINE and CRISTAL observations were combined and re-processed by the CRISTAL team consistently.
Several data products, including [CII] moment maps and dust continuum emission maps, each with Briggs (robust$=0.5$) and natural weighting schemes, were generated.

In this work, we adopt the natural-weighted data products for their higher signal-to-noise ratio (SNR). For these, the median synthesised beam size for our subsample of 14 sources is $0.45\arcsec\times0.54\arcsec$. To enable our multi-wavelength analysis, we converted the dust continuum images from Jy/beam to Jy/pixel to match with the HST and JWST images. The equivalent area of the Gaussian beam ($\Omega$) was calculated using the equation, \begin{equation}
    \Omega = \frac{\pi\theta_\mathrm{\rm maj}\theta_\mathrm{\rm min}}{\rm 4\ln{2}}\,,
\end{equation}
where $\theta_\mathrm{maj}$ and $\theta_\mathrm{min}$ are the FWHM of the ALMA synthesized beam in pixel units along major and minor axis, respectively \citep{Meyer2017}.

%{\bf Say something about the typical depth/sensitivity of these observations. What is the typical depth in the continuum?}
% \begin{table}
%     \centering
%     \begin{tabular}{c|c|c|c}
%         CRISTAL ID & BMAJ(\arcsec) & BMIN(\arcsec) & $\rm F_{\rm 850\mu m}(mJy)$\\
%          C02 & 0.54 & 0.48
%          C03 & 0.77 & 0.65 & $0.045\pm0.031$\\
%          C04 & 0.76 & 0.58 & $0.26\pm0.12$\\
%          C06 & 0.54 & 0.44 & $0.37\pm0.13$\\
%          C07 & 0.72 & 0.47
%          C09 & 0.34 & 0.31 & $0.17\pm0.067$\\
%          C11 & 0.50 & 0.39 & $0.22\pm0.093$\\
%          C13 & 0.54 & 0.45 & $0.058\pm0.058$\\
%          C14* & 0.12 & 0.11 & $0.65\pm0.20$\\
%          C19 & 0.66 & 0.53 & $0.084\pm0.033$\\
%          C21 & 0.36 & 0.32 & $0.22\pm0.086$\\
%          C24 & 0.30 & 0.24
%          C25 & 0.29 & 0.24
%     \end{tabular}
%     \caption{Caption}
%     \label{tab:my_label}
% \end{table}
\subsection{HST/ACS and WFC3 observations}\label{sec:hst}
We used the {\it grizli} pipeline \citep{brammer2023} to retrieve and reduce the archival HST/ACS and HST/WFC3 observations for all our selected targets.
% The pipeline pulls all raw exposures from MAST archive that overlaps with the set of coordinates of our targets.
% It then perform automatic data reduction procedures to generate final mosaics.
For each filter, the pipeline retrieves, calibrates and re-samples all available individual raw exposures that overlap with the target coordinates.
Then, the calibrated frames were carefully aligned using various astrometric reference catalogs including the DESI Legacy Imaging Survey DR9, PANSTARRS(PS1), and {\it Gaia} (we return to the topic of astrometric alignment in Section~\ref{align}).
Lastly, they were combined into the final mosaics.
% Hence, the output would be the deepest ever HST observations of the CRISTAL galaxies.

% {\bf Justin Spilker} reduced the existing HST observations with grizli. Four filters: F814W, F105W, F125W, F160W (60mas).
% {\bf So grizli automatically pulls all raw exposures that overlap with a given set of coordinates, along with some extra buffer space in case you also want to process a slightly wider area around a target. so in theory the images from before should have all available exposures reduced jointly. }

\subsection{JWST/NIRCam observations}\label{sec:jwst}
All sources in our sample have been observed with at least 4, and up to 8, JWST/NIRCam filters as part of the PRIMER (PID: 1837; PI: Dunlop), COSMOS-Web (PID 1727; co-PIs: Kartaltepe \& Casey; \citealt{Casey2023}) \textcolor{black}{and a NIRCam parallel  (PID 3990; PI: Morishita)} GO programs. The raw data are publicly available from MAST archive. 

For the sources in the PRIMER-COSMOS field (i.e., CRISTAL-11 and CRISTAL-13), we used the reduced data products from the Dawn JWST Archive (DJA) Mosaic release v7 \citep{Heintz2024}, which used the \texttt{grizli} software \citep{brammer2023}.
The general data reduction processes are described in \citet{Valentino2023ApJ}.

For the remaining sources, which are in the COSMOS-Web field, we reduced the data using the standard JWST Calibration python package \textcolor{black}{(version=1.10.0, pmap=1075, \citealt{jwst_pipeline_1.10.0})} with highly optimized parameters using the {\footnotesize CRAB.Toolkit.JWST}\footnote{\url{https://github.com/1054/Crab.Toolkit.JWST}} package.
This package contains image processing tools other than the standard pipeline which were applied in this work, including (1) 1-over-f noise removal using the method described in \citet{Bagley2023ApJ}; (2) background removal via the skymatch method in the standard pipeline; (3) known artifacts removal with wisps template following \citet{Bagley2023ApJ} and manual masking for varying artifacts like claws; and (4) astrometric alignment with {\it Tweakreg} to align to \textcolor{black}{ the COSMOS2020 catalog that incorporated the Gaia-DR3 astrometry}.

\textcolor{black}{We have inspected the images reduced from both our pipeline and the DJA mosaics to verify they are consistent. We generated photometric catalogs on the PRIMER-COSMOS field where we have data products from both methods, the measured magnitudes from the two catalogs are consistent within 0.1 magnitude for sources brighter than 26 in AB magnitude in all 8 NIRCam filters used in this work.}

\section{Method}\label{sec:method}
\subsection{Image Processing}
To properly combine the multi-wavelength information from HST, JWST, and ALMA in a resolved manner, we need to further process the data because the astrometry and angular resolution may differ from each other.
Here, we describe the steps to obtain correct absolute and relative astrometric alignments, and homogenize the point-spread function (PSF).

\subsubsection{Absolute Alignment}\label{align}
Images from each instrument and filter were independently calibrated for their astrometric alignment.
HST images were aligned during the {\it grizli} reduction pipeline. JWST images were also aligned using either {\it grizli} or {\it Tweakreg}. 
ALMA observations were aligned to the International Celestial Reference Frame (ICRF) via compact calibrators and should be the most accurate among the three telescopes. Nonetheless, there can still be minor misalignments between different filters, as they were independently calibrated and each has different astrometric accuracy. This is relevant because even small misaligments may have an impact on our spatially resolved SED analysis by increasing/decreasing color gradients artificially if our spatial bins are smaller than or similar to these potential offsets.

To ensure that each spatial bin corresponds to the same area of sky, we verified the relative alignment between all filter images.
First, we re-sampled all images to the same pixel grid with size of 0.03\arcsec/pixel using {\it ProPane} \citep{Robotham2024}. \textcolor{black}{We carefully inspected that the resampling of already drizzled mosaics do not induce any bad artifacts. The resampling, however, increased the correlation among nearby pixels which are smaller or comparable to the typical sizes of the pixel binning as discuss later in Sect.~\ref{sec:apgrid}.}

Then, we created source catalogs for all the HST images and computed the median shifts in pixel units, if any, of the source locations for each image relative to the HST/ACS F814W filter.
We choose F814W as a reference because of its smaller pixel area, smaller PSF size and closer wavelength to {\em Gaia} observations. These factors should make it more accurately aligned to the World Coordinate System(WCS).
Note that, when correcting for any relative misalignment, we apply only linear shifts.

After all the HST images were aligned, we aligned all the JWST images in a similar manner, using the aligned HST/WFC3 F160W image as reference.
\textcolor{black}{We choose F160W because it allows more sources to be matched between HST and JWST filters}, and hence increases the robustness of our alignment correction.

For ALMA observations, due to the small field of view and much longer wavelength, it is impossible to check for misalignments relative to HST and JWST using the source catalog method. We have to rely on the absolute astrometric calibration of this telescope. The nominal pointing accuracy of ALMA observations is given by $\rm pos_{acc} = beam_{FWHM}/SNR/0.9$ (see section~10.5.2 of the ALMA Technical Handbook\footnote{\url{https://almascience.eso.org/documents-and-tools/cycle11/alma-technical-handbook}}). For observations with angular resolution of $0.5\arcsec$ and $\mathrm{SNR}\geq10$, the nominal astrometric accuracy would be $\lesssim 0.06\arcsec$, which is smaller than the spatial scale of our resolved study\footnote{We note that all ALMA observations used here were of sufficiently long duration that any astrometric shifts caused by a temporary atmosphere phase screen would be averaged out.}.

We note that all the relative misalignments mentioned above were sub-dominant even before any corrections. Typical offsets were measured to be at sub-pixel scale ($<0.03\arcsec$) and at most no more than 3 pixels ($<0.09\arcsec$). These small offsets are negligible comparing to the typical PSF size $\sim0.5\arcsec$ (i.e. ALMA beam FWHM at major axis), as well as the sizes of the spatial bins used later in our analysis ($0.12\arcsec-0.15\arcsec$). Hence, we conclude that the uncertainty in astrometric alignment would not affect our spatially-resolved analysis based on the spatial scales adopted in this work.

% \subsubsection{Flux conversion}
\subsubsection{PSF modeling and homogenization}
Before moving on to SED fitting, we match the point spread functions (PSFs) of all the observations to avoid potential artificial color gradients caused by optical effects. 
To do that, PSF models are required for all the filters covering each target field.

For the HST/WFC3 filters, we use the {\it grizli} PSF tool and the effective PSF models\footnote{\href{https://www.stsci.edu/hst/instrumentation/wfc3/data-analysis/psf}{https://www.stsci.edu/hst/instrumentation/wfc3/data-analysis/psf}}.
The tool drizzles the model PSFs of each individual exposure with the same drizzle parameters that were used for the science images.
For HST/ACS F814W, which is the only ACS filter used in this work, we use TinyTim \citep{Krist2011SPIE} to create the PSF model for each CRISTAL galaxy following Cook et al.~(subm.). 
For a given galaxy, we first simulated a number of PSF models using TinyTim, with parameters depending on where the targeted galaxy landed on the detector in each exposure, to best capture the focal plane variation. 
% This is required because the COSMOS field has been visited by HST multiple times, and the PSF may change significantly among different visits.
We then rotated these model PSFs of individual exposures according to the position angle of the observations, and combined them by weight-averaging with exposure time.
\textcolor{black}{We note that the obtained PSF models are only approximations to the true PSF due to various technical challenges.}
However, we deem these F814W PSF models to be adequate for our purposes, as the final targeted PSF (i.e., the ALMA beam) is much larger than the F814W PSF, hence, slightly different F814W PSF models would be indistinguishable after convolving to the targeted resolution.

We model the JWST PSF using the {\it WebbPSF} package \citep{Perrin2014SPIE}. 
The {\it WebbPSF} package transforms optical path difference (OPD) maps measured in-flight into PSFs, taking into account detector pixel scales, rotations, filter profiles, etc. 
We generated a PSF model for each galaxy and each filter sampling at the $0.03\arcsec$/pixel detector pixel scale, using the OPD of the date of observation ({\sc date-obs}), and rotated it using the position angle of JWST ({\sc pa\_aper}) stated in the image header.
Some CRISTAL galaxies in the PRIMER-COSMOS field were imaged in multiple adjacent observations. However, they were taken consecutively within roughly two weeks. Their relative rotation is negligible ($<1^{\circ}$), the same exposure time and dithering pattern were used, and the OPD is usually stable over short time periods. Thus, we used the same averaged {\sc date-obs} and {\sc pa\_aper} for all observations for simplicity. % (i.e. 2022-12-28 and 291.5$^{\circ}$).

\textcolor{black}{We noted that recent studies found the {\it WebbPSF} simulated PSF tend to be slightly sharper than empirically created PSF (ePSF) especially in shorter wavelength filters \citep{Weaver2024}. We also found similar difference between ePSF and simulated PSF but we are cautious about the accuracy of the ePSF as we found that it is sensitive to how well one can determine the center of stars used to create the ePSF, which can be uncertain for short wavelength filters as the FWHM of stars are often smaller than 2 physical pixels  (i.e. under sampled). Sharper PSF models at short wavelength would result in over-spreading of optical blue/UV light after matching to ALMA resolution. As we find in Sect.~\ref{sec:fobs-trend} and discuss in Sect.~\ref{sec:fobs-r}, dust emission seems to be surprisingly more extended than the UV emission even when we adopt a sharper PSF model for the UV filter. 
Therefore, we decide to adopt the simulated PSF because it is less method-dependant and more conservative.}

For ALMA, we generate 2-dimensional elliptical Gaussian profiles using the information provided in the header, as the synthesised beam of the observations is also the target PSF for convolution during PSF homogenisation.

\begin{figure}
    \centering
    \includegraphics[width=0.45\linewidth]{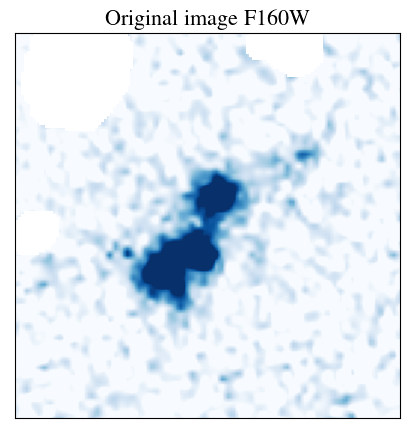}
    \includegraphics[width=0.45\linewidth]{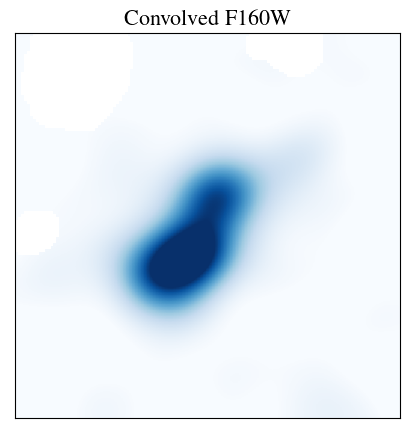}
    \includegraphics[width=0.45\linewidth]{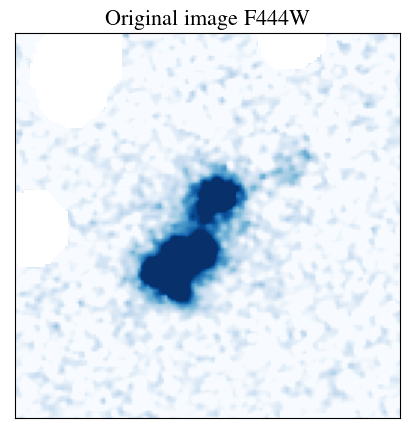}
    \includegraphics[width=0.45\linewidth]{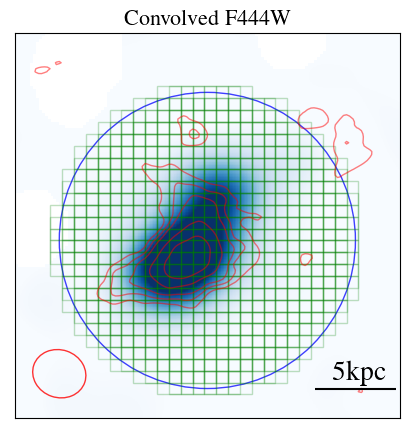}
    \caption{Comparison between the native resolution images of one of our galaxies, CRISTAL-02 (left-hand panels), and the corresponding convolved images resulting from our PSF homogenization procedure (right-hand panels). The top panels show the images in filter HST/WFC3 F160W, and the bottom panels show the images in filter JWST/NIRCam F444W. In the bottom-right panel, we overlay the ALMA Band 7 continuum emission as red contours (showing steps of 2-, 3-, 4-, 6-, 8-$\sigma$; the synthesized beam FWHM size shown in the bottom-right corner). The blue circle has a diameter of $3\arcsec$ and indicates the approximate size of the total aperture for this source. The grid of green square boxes of $0.12\arcsec$ on each side indicates the grid of spatial bins where fluxes are extracted to model the SED.}
    \label{fig:aperture_grid}
\end{figure}

We then create matching kernels between PSF models of each filter to the target PSF (i.e., the ALMA synthesised beam) using the {\it create\_matching\_kernel} function in {\it Photutils} package \citep{photutils2022}, with a low-pass window function to minimize spectral leakage.
Then, we convolve all the images with the matching kernels using the {\it convolve} function in {\it Astropy} \citep{astropy:2022}. In Fig.~\ref{fig:aperture_grid}, we show images of one of our sources, CRISTAL-02, in the HST/WFC3 F160W and JWST/NIRCam F444W filters, before and after our PSF homogeneisation procedure.

\subsection{Spatially-resolved SED fitting}

\subsubsection{Aperture grid photometry}
\label{sec:apgrid}
To perform spatially-resolved SED modeling, we place apertures at different sightlines of the galaxy; then, we extract the fluxes and model the SED for each aperture.
We applied a uniform square tessellation (hereafter, the aperture grid) that fully captures all the fluxes in all filter bands as a whole, and yet uniformly resolves the galaxy regardless of potential variations in surface brightness and morphology across the wide wavelength range of our observations.
Comparing to an adaptive scheme that aims to unify the significance (e.g., overall SNR, total luminosity) in each aperture, our spatially uniform grid simply unifies the spatial scale.
The main reason for this choice is that our galaxies are not very extended compared to the ALMA beam size, and therefore we cannot afford more sophisticated aperture grids for these sources. Furthermore, adaptive apertures might bias towards certain features or physical components depending on the method used. In particular, if we used a uniform SNR method, the grid would focus on (young) stars and/or dust-free regions, as the HST and JWST observations generally have higher SNR than ALMA. If we used a uniform luminosity method, the grid would focus instead on the dust emission.
Additionally, given the potential physical offsets between UV and IR emission peaks (e.g., \citealt{Inami2022}), a uniform aperture grid is best suited for wide wavelength range studies.

As an example, we show the aperture grid for CRISTAL-02 in Fig.~\ref{fig:aperture_grid} (bottom-right panel).
The square boxes in the aperture grid serve as the spatial bins of our resolved study. The circle defines the boundary we manually chose for each source to capture the whole galaxy as viewed in all HST and JWST bands after resolution-matching to ALMA. We add up fluxes of all spatial bins on or within this circle to obtained the total, integrated fluxes.

We adapt the spatial bin sizes to the ALMA beam size: they are $0.15\arcsec$ on each side if both the major and minor ALMA beam axes are larger than $0.5\arcsec$; $0.09\arcsec$ on each side if both the major and minor ALMA beam axes are smaller than $0.3\arcsec$; and $0.12\arcsec$ a side for all other cases. 
This ensures that we consistently resolve the PSF with $\simeq4$ spatial bins along both major and minor axes for all galaxies in our sample.
Further reducing the spatial bin size would not provide additional meaningful spatial information due to the relatively coarse PSF. Moreover, smaller bin sizes would increase the errors per bin, and make the study more sensitive to any subtle astrometric misalignment among different filters/instruments as discussed in Section~\ref{align}.
At the redshifts of our galaxies, the spatial bins correspond to physical sizes from $\simeq0.5$ to $\lesssim1$\,kpc.

After defining the aperture grid, the next step is to perform resolved photometry by simply adding up the fluxes in all the pixels in each square spatial bin.
The flux uncertainties for each filter are sampled directly from the images.
We randomly place square apertures of the same size in the blank area of the masked image, and measure the total flux inside the aperture.
We repeat this process 1000 times to obtain a sky flux distribution measured with this aperture; we then compute the 16th and 84th percentiles of the sky flux distribution (after applying a 5$\sigma$ clipping to remove contamination from imperfect masking, if any). The flux uncertainty is finally obtained by halving the difference between 16th and 84th percentiles.
%We also measure the total integrated photometry by adding up the fluxes in all the apertures for all observed filters. 
%The uncertainty of the total photometry is also sampled in the masked image in a similar manner using the integrated aperture (i.e. collection of all spatial bins).

The filter coverage of our sources is not uniform, however all of them are observed at least in the following set of seven filters: HST F814W and F160W, JWST F115W, F150W, F277W, and F444W, and ALMA Band 7. Furthermore, we use HST F140W observations available for eight sources,  HST F105W for seven sources, HST F110W and F140W for four sources, and JWST F200W and F356W for three sources, and JWST F090W, F277W, and F410M for two sources.

\subsubsection{SED fitting with \magphys}
The previous step allows us to extract the observed spectral energy distribution (SED) from the region indicated by green boxes in Fig.~\ref{fig:aperture_grid}, i.e., each spatial bin.
We do not require each spatial bin to have detected signal in all filters to avoid introducing biases, since different wavelengths trace different physical components of the galaxy that could have spatial offsets in projection. 
SEDs extracted in any aperture with more than $1\sigma$ flux detection (i.e., SNR$>1$) in at least four filters were modeled with \magphys\ \citep {magphys2008,magphys2015}. When there is no detection in a given filter, we use a \textcolor{black}{$3\sigma$ upper limit}, which still provides meaningful constraints in \magphys\ by penalizing models with flux predictions that exceed the upper limit.
As there may exist regions in the galaxy that are detected only in the optical/near-IR or far-IR but not both, we adopted this approach to include all the fluxes in any filters into our analysis. 
We have tested different combinations of number of filters above certain flux levels (e.g., 3 filters above $2\sigma$ or 5 filters above $1\sigma$) and found that they all capture most of the galaxy light with slight variations in low surface brightness regions.
The caveat of doing this is that some SEDs may be loosely constrained when there are multiple upper limits. These low surface brightness regions only appear at the outskirts of the galaxies. We find that these regions do not contribute significantly to the total inferred physical quantities, and they do not affect the analysis due to their larger error bars.
% Note that despite each galaxy in our sample may be observed with different number of filters, from 3 to 5 HST filters and from 4 to 10 JWST filters, we consistently use 4 filters for spatial bins selection criteria.  
% {\bf How do \magphys handles upper limit?}

{\sc magphys} employs an energy balance technique to model the dust attenuation of stellar emission in the UV/optical/near-IR, and the re-emission of this energy in the far-IR by dust grains in a consistent way. The code uses the stellar population synthesis models of \cite{Bruzual2003}, with a \cite{Chabrier2003} IMF between $0.1$ and $100\,M_\odot$, and randomly samples stellar metallicities in a flat prior between $0.2$ and $2\,Z_{\odot}$. The star formation history (SFH) is parametrised as a continuous delayed exponential function with varying timescales and includes superimposed random bursts to account for starburst stochasticity (see \citealt{magphys2015} for details). Dust attenuation is modeled using the two-component of \cite{Charlot2000}, which distributes interstellar dust in star-forming regions (stellar birth clouds) and in the diffuse interstellar medium (ISM). This results in an age-dependent dust attenuation of starlight, wherein young stars (ages $<10$\, Myr) are attenuated by dust in the birth clouds and in the diffuse ISM, and older stars are only attenuated by dust in the diffuse ISM. Here, we use the refinement of this model by \cite{magphys2020}, which includes a variable $2175$\,\AA\ feature in the dust attenuation curve.
The dust emission is modeled using an empirical \textcolor{black}{polycyclic aromatic hydrocarbon (PAH)} emission template and a set of modified black bodies with varying dust temperatures, i.e., two hot components with $T_{\rm dust}=130$\,K and $250$\,K that reproduce the mid-IR continuum, and two cooler components with $T_{\rm dust}=30-80$\,K (with dust emissivity index $\beta=1.5$) and $T_{\rm dust}=20-40$\,K (with $\beta=2$); these result in a prior on the luminosity-weighted dust temperature that ranges from 20 to 80\,K. More details on the parameters and priors can be found in \citet{magphys2015}. 

The \magphys\ code uses a Bayesian approach to fit the SEDs. First, it generates a large library of models where each free parameter is stochastically sampled from its prior distribution. Then, each model in this library is compared to the observations (in this case, the observed SED of each spatial bin) via the $\chi^2$ goodness-of-fit, which obtains the probability of each model matching the observations. The posterior likelihood distribution of each parameter is then obtained by marginalizing over the probabilities of all other parameters. In this study, we focus on the following output parameters: stellar mass ($M_\ast$), star formation rate averaged over the last 100 Myr (SFR), average $V$-band dust attenuation (\av), and total dust luminosity (\ldust; integrated between 3 and 1000\,$\mu$m). The Bayesian fitting method ensures that uncertainties due to flux errors, incomplete SED sampling, and intrinsic parameter degeneracies are all included in the posterior likelihoods.

\section{Results}\label{sec:results}
\subsection{\magphys\ parameter maps}

% In general, the SED fitting results were acceptable at where the SNR is high (i.e. the inner part of the galaxies). There may be cases that some resolved SEDs near the edge are not well fitted (i.e. $\chi^2 \gtrsim 2$), the inferred parameters would hence have relatively larger uncertainties.
Following the matching and homogenization of our images described in the previous sections, we are left with a total of 2,730 spatial bin SEDs to model with \magphys\ across our sample; on average, each galaxy consists of 195 spatial bins. We note that, although adjacent bins can be correlated because of our PSF smoothing, each spatial bin is modeled independently by \magphys, that is, no information about adjacent spatial bins is taken into account during the SED modeling.

The resolved SEDs of all CRISTAL galaxies are generally well fitted, \textcolor{black}{the $\chi^2_{\mu}$ (normalized by the number of filters) of all best-fit resolved SEDs of each galaxy} follows a log-normal distribution peaking at \textcolor{black}{$0.1<\chi^2_{\mu,peak}<1$, showing that majority of the spatial bins can be well described by the \magphys\ models within flux uncertainties}. A small number of SEDs are more poorly fitted \textcolor{black}{($\sim1.5$\% of the total)}, with a $\chi^2_{\mu}$ up to $\sim2$. This usually happens in galaxy outskirts, where both the surface brightness and number of detected filters are low. The only exception to this is CRISTAL-25, for which we found the central region to be poorly fitted. Further investigation reveals this is caused by a flux excess in F150W, F160W (rest-frame $\lambda\simeq2800$\AA), and F277W (rest-frame $\lambda\simeq5000$\AA), which could be hints of AGN emission line contamination. We discuss this in more detail in Appendix~\ref{app:C25}.

\begin{figure*}[ht!]
    \centering
    \includegraphics[width=\linewidth]{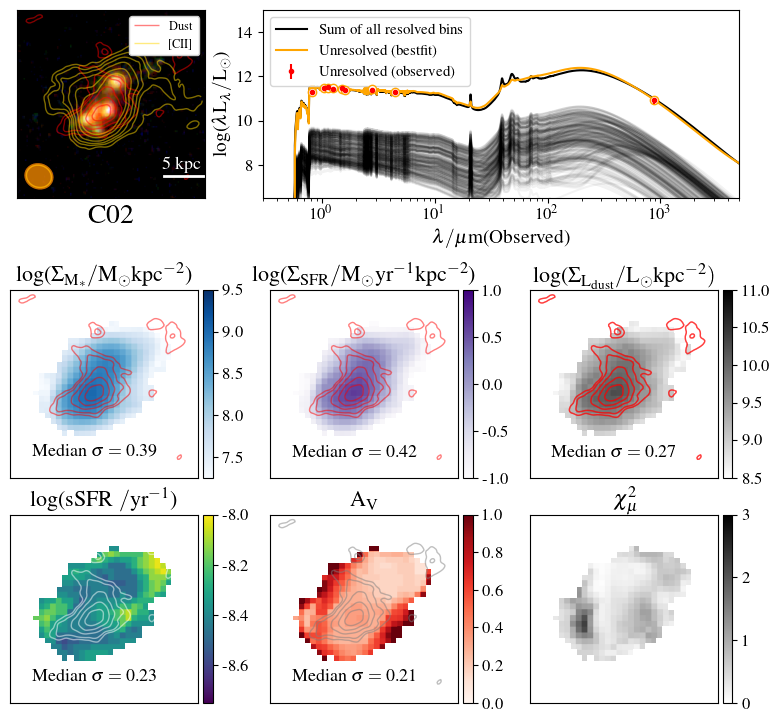}
    \caption{Physical parameter maps of CRISTAL-02 obtained with \magphys. The RGB image in the top left-corner is created using three native resolution JWST/NIRCam filters: F150W, F277W and F444W as B, G and R respectively, with the scaling following the prescription from \citet{Lupton2004}. The overlaid red contours show the ALMA Band 7 dust continuum emission at 2-, 3-, 5-, 7- and 9-$\sigma$. The same dust contours are also plotted in other panels. The yellow contours show the ALMA [CII] emission at 3-, 5-, 7-, 10-, 15-, and 20-$\sigma$. The ALMA synthesised beams of these observations are shown in the bottom-left corner. A 5 kpc scale bar is shown at the bottom-right; the scale is the same in the bottom panels. In the top-right panel, each grey curve is the best-fit SED model for each spatial bin (i.e., each pixel in the parameter maps). The sum of all best-fit model curves is shown as the solid black curve. The red points are the total integrated photometry of the galaxy in the available HST, JWST and ALMA bands, and the orange line shows the best-fit \magphys\ model to those points.
    The middle panels show, from left to right, the spatial distribution of stellar mass surface density ($\Sigma_{M_\ast}$), star formation rate surface density ($\Sigma_{\rm SFR}$), and dust luminosity surface density ($\Sigma_{L_{\rm dust}}$); the bottom panels show the spatial distribution of specific star formation rate (sSFR), average $V$-band dust attenuation ($A_{\rm V}$), and $\chi^2_{\mu}$. The median uncertainties of the parameters are shown in each panel.
    (the complete figure set (14 images) is available in the online journal.)}
    \label{fig:maps}
\end{figure*}

In Fig.~\ref{fig:maps}, we present an example of our resolved analysis of CRISTAL-02, the most massive and and the most star-forming galaxy in the main CRISTAL sample. It also has angular resolution (i.e., ALMA beam size) similar to the median of the whole sample.
The top-left panel shows a RGB image composed with the native resolution (i.e., before convolution) NIRCam images (rest-frame UV and optical), with the ALMA dust continuum and [CII] line emission overlaid as red and yellow contours, respectively. The ellipses in the lower left corner indicate the ALMA beam sizes and show that both the dust and [CII] emission are spatially resolved.
We notice multiple visual `clumps' resolved by JWST within the proximity of the dust and [CII] emissions. However, the sizes of these `clumps' and their separations are small compared to the ALMA beam size. It is hard to differentiate whether these `clumps' are physically separated or if they only appear like `clumps' because of dust extinction. 
% The UV/optical color appears to be redder in between the `clumps' where the dust emission peaks, and more greenish or bluish further out where the dust emission is lower.
% In some other CRISTAL galaxies with more separated `clumps', the dust emission more obviously peaks between the optically-bright `clumps', which seems to favour the dust obscuration scenario.
%
%Fitting the full spatially-resolved SEDs  with \magphys\ should recover physical parameter distributions that take into account the degenerate effects of dust obscuration and age in a self-consistent manner.

The top-right panel of Fig.~\ref{fig:maps} shows the best-fitting \magphys\ SEDs of all the spatial bins in this source, as well as their sum, and compares the sum to a fit of the total integrated SED. The sum of the best-fit models of the spatial bins and the best-fit to the total integrated SED agree extremely well through the whole UV to far-IR range, even in the near- to far-IR where we have little to no observational constraints. This shows that our spatially-resolved approach robustly recovers the integrated properties.

The six bottom panels in Fig.~\ref{fig:maps} show the spatial distributions of some of the physical parameters constrained by \magphys\ (stellar mass, star formation rate, dust luminosity, specific star formation rate, and average $V$-band dust attenuation), as well as the \textcolor{black}{`fiducial'} $\chi^2_{\mu}$, i.e., the $\chi^2$ divided by the number of filters available. For each parameter map, we also indicate the median uncertainty for each parameter, which is the median of the uncertainties in all spatial bins, given by half of the 16th--84th percentile of the posterior distributions (e.g., \citealt{magphys2008}).
We find that the stellar mass, SFR and dust luminosity parameter maps of CRISTAL-02 are smooth, with no signs of clumpiness and other morphological features. However, we cannot disentangle the effect of PSF-matching to the worst resolution of our observations (in this case, ALMA). 
The stellar mass and SFR surface densities seem to correlate well spatially (we further analyze this for the whole sample in Section~\ref{sec:rSFMS}).
The dust luminosity surface density peaks where the ALMA continuum emission peaks, as expected. 
We note that this parameter is relatively loosely constrained, as it depends on the dust temperature, and we lack ALMA observations sampling the peak of the dust continuum emission (e.g., \citealt{daCunha2021}). However, the energy balance done by \magphys\ helps constrain this parameter using additional information from HST and JWST \textcolor{black}{because the two components in the dust attenuation models are separately tied to cold and warm dust emission components}.
The specific star formation rate (sSFR) map shows more structure. The sSFR is often used to probe past versus present star formation, and its radial variation can be used to test inside-out or outside-in growth scenarios (e.g., \citealt{Nelson2021}). The sSFR map of CRISTAL-02 shows multiple local maxima that are slightly higher ($\lesssim0.5$ dex) than their surroundings; however, we observe no clear radial trends at the resolution of our maps. The peaks in sSFR maps may be driven by the different morphology between stellar mass and ALMA emission which highlight small regions of enhanced recent (obscured) star formation activity.
The distribution of $V$-band dust attenuation, $A_\mathrm{V}$, appears relatively smooth and flat, and tends to coincide with bright ALMA emission and red UV/optical colors. In some cases, we observe an increase in $A_\mathrm{V}$ towards the outskirts, which coincides with extended ALMA emission, though this may be less reliable due to the lower SNR at large radii \textcolor{black}{and the MAGPHYS prior on Av being relatively high for `typical' star-forming galaxies (peaking at \av$\sim2$, \citealt{magphys2015})}.

\subsection{Resolved star-forming main sequence}\label{sec:rSFMS}

\begin{figure*}
    \centering
    \includegraphics[width=\linewidth]{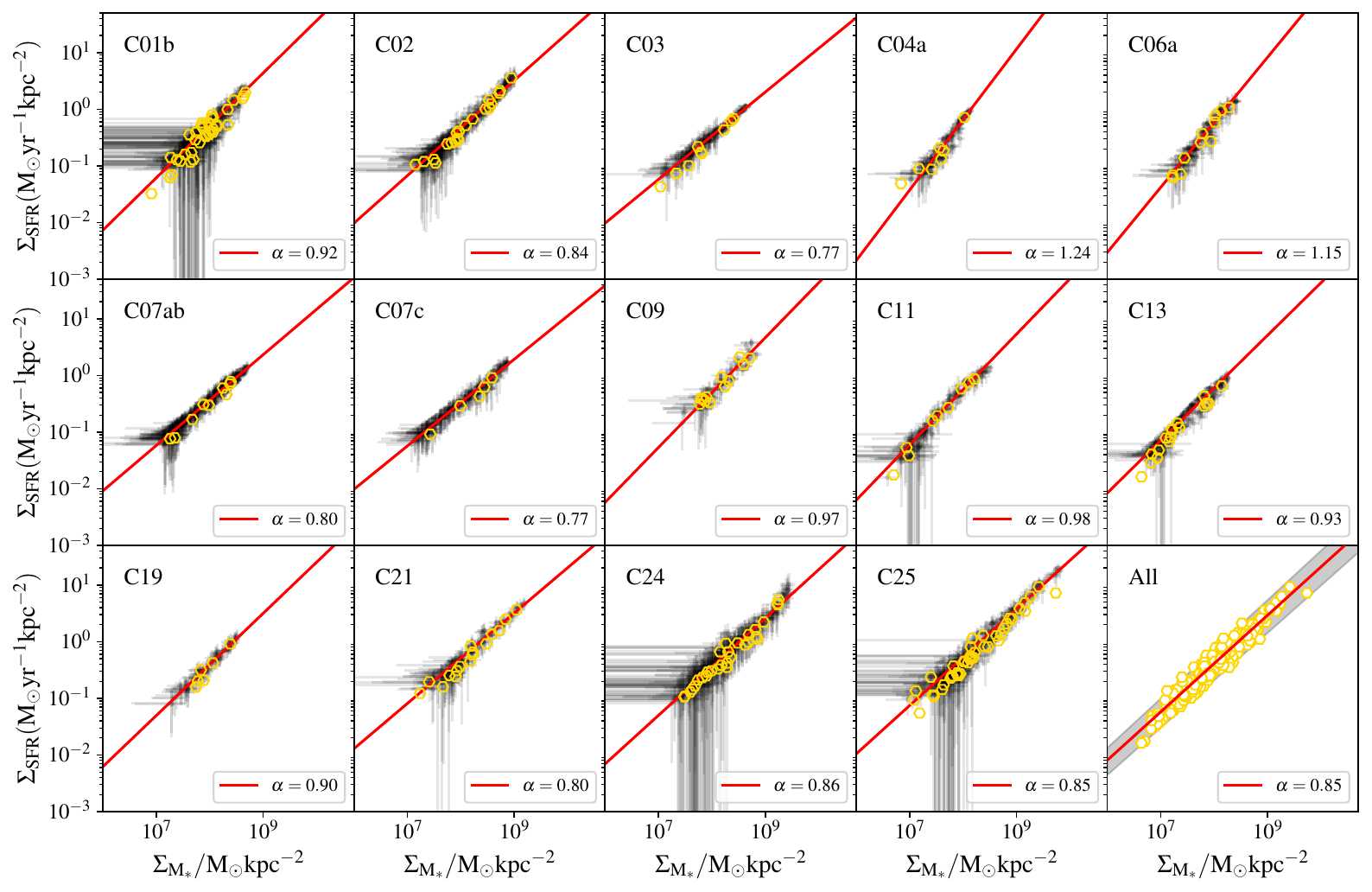}
    \caption{The spatially resolved star-forming main-sequence (rSFMS) of our CRISTAL galaxies. Each panel (except the last) shows the surface density of SFR ($\Sigma_{\rm SFR}$) plotted against surface density of stellar mass ($\Sigma_{\rm M_*}$) for an individual galaxy (ID on the top-left corner). The black crosses correspond to our spatial bins ($\sim0.5-1$\,kpc). For each spatial bin, the values are derived from the median-likelihood \magphys\ estimates using the full UV-to-IR SED fitting, and the error bars are based on the 16th-84th percentiles of the posterior likelihood distributions.
    The red lines show the best-fit power laws; for each one, we indicate the power-law slope $\alpha$ in the bottom-right corner. 
    The yellow hexagons correspond to using a hexagonal tessellation with spatial bin sizes comparable to the ALMA beam, such that adjacent bins are less likely to be correlated due to flux blending.
    The last panel shows the joint distribution of the whole sample (hexagon bins only), which we use to measure the best-fit power law slope to the rSFMS at $z\simeq5$.}
    \label{fig:rSFMS}
\end{figure*}

The integrated star formation rates of galaxies and their stellar masses have been shown to correlate tightly from the local Universe to high redshifts, a relation known as the `star-forming main sequence' (e.g., \citealt{Brinchmann2004,Noeske2007,Daddi2007,Whitaker2012,Speagle2014,Popesso2019,Villanueva2024a}). This empirical relation has also been explored in resolved studies of local galaxies, which relate the surface density of SFR  ($\Sigma_{\rm SFR}$) with the surface density of stellar mass ($\Sigma_{\rm M_*}$) on $\sim$kpc scales and better, finding a tight `resolved star-forming main sequence' (rSFMS; e.g., \citealt{Cano-Diaz2016,Abdurrouf2017,Maragkoudakis2017,Liu2018,Ellison2021,pessa2021,Sanchez2021,Mun2024,Koller2024}). There is some indication that a rSFMS can be found in galaxies at least until $z\simeq2$ (e.g., \citealt{Wuyts2013}). The physical origin of this rSFMS is still debated. \cite{Abdurrouf2017} find that it is a sign of spatially constant specific star formation rates through galaxy disks. \cite{Baker2022} argue that it is not an intrinsic correlation, but rather a result of two more fundamental relations: the resolved Kennicutt-Schmidt law, i.e., the correlation between SFR and gas surface densities, and the resolved molecular gas main sequence, i.e., the correlation between the surface densities of stellar mass and molecular gas (see also, e.g., \citealt{Lin2019,Ellison2021,Saintonge2022,Villanueva2024a}). Nevertheless, the slope of the rSFMS on $\sim$kpc scales has been linked to galaxy morphologies (e.g., \citealt{Maragkoudakis2017}) and to different assembly modes in galaxies (e.g., \citealt{Liu2018}).

Here we use our star formation and stellar mass maps to investigate if a rSFMS is present in our galaxies.
We plot \sigsfr\ against \sigmstar\ for all our CRISTAL sources in Fig.~\ref{fig:rSFMS}.
The black points show the median-likelihood parameters for each spatial bin given by \magphys. As described in Section~\ref{sec:apgrid}, these spatial bins correspond to physical sizes between 500~pc and 1~kpc at the redshifts of our sources.
We find that all CRISTAL galaxies show strong correlation between \sigmstar\ and \sigsfr, i.e., we retrieve a resolved star-forming main sequence at $z\simeq5$.
\textcolor{black}{We quantify this relation by fitting a power law function, $\sigsfr \sim \sigmstar^\alpha$, for each source. The power-law slopes for our sources range from $\alpha=0.77$ to $\alpha=1.24$. The steepest relations are seen in sources CRISTAL-04a and CRISTAL-06a, with $\alpha=1.24$ and $\alpha=1.15$, respectively. These happen to be sources with close nearby companions (CRISTAL-04b and CRISTAL-06b, which are not dust detected), and they are also the two sources with the most centrally peaked specific SFRs (See the complete figure set of all maps available in the online journal). Perhaps interactions in these systems lead to an increase in star formation efficiency, which would result in the steeper slopes.}

We note that the black points in Fig.~\ref{fig:rSFMS} are not independent to each other, as we expect nearby spatial bins to be correlated due to flux blending. To test if this affects our results, we use larger resolved apertures (using a uniform hexagonal tessellation over the same region) such that the spatial bins are comparable to or larger than the ALMA beam in all of our targets ($\sim0.5$\,arcsec, which corresponds to physical sizes $\sim3$\,kpc at the redshifts of our sources). We plot the results as yellow hexagons on Fig.~\ref{fig:rSFMS}. The rSFMS remains unchanged, confirming that the correlation between surface SFR density and the surface stellar mass density is genuine, and independent of the spatial bin size down to the scales probed by our observations. 

\textcolor{black}{We verified the results by fitting two of our sample galaxies (i.e. C11 and C13 which have 4 HST and 8 JWST filters) with a different SED fitting code \texttt{ProSpect} \citep{Robotham2020}, adopting a non-parametric SFH with only the HST and JWST observations. We also obtained rSFMS with power law slope comparable with \magphys\ results but found systematic offset in inferred stellar mass and larger scatter in the obtained rSFMS. The larger scatter can be the consequence of a more flexible SFH model in \texttt{ProSpect} which inferred more dispersed ages with larger uncertainty than \magphys. The offset is likely due to that \texttt{ProSpect} typically finding older ages than \magphys\ \citep{Leja2019,Bellstedt2020,Thorne2021}. The lower scatter in the rSFMS from \magphys\ may be caused by its parametric SFH model, which favours younger average stellar ages relative to non-parametric SFHs in the absence of age-constraining information. This may result in a lower dispersion in the stellar masses and SFRs relative to other SED modelling codes. }
% We do not have the ground truth on whether different parts of a galaxy at z$>4$ would have very different ages, so it is not clear that whether one is better than the other.
% Though we are not sure to what extent different parts of a galaxy would be dispersed in ages at z$>4$ when the Universe is younger than $\sim1.5$Gyr.

In the bottom right-hand panel of Fig.~\ref{fig:rSFMS}, we combine the whole sample (hexagon bins only). We find a best-fit power law slope of $0.85\pm0.02$, with a median scatter of $\sim0.1$\,dex. This is close to the slope found by \cite{Wuyts2013} in $0.7<z<1.5$ galaxies, who find a rSFMS slope of 0.84 on $\sim$\,kpc scales, and broadly consistent with resolved studies of low-redshift galaxies at similar $\sim$kpc scales, where slopes between $\sim0.7$ and $\sim1$ are found (see compilation by \citealt{pessa2021}).
Our results extend previous findings in redshift and indicate that this sub-galactic scaling relation persists up to $z\simeq5$.

\subsection{Resolved dust obscuration}\label{sec:fobs-trend}

\begin{figure*}
    \centering
    \includegraphics[width=\linewidth]{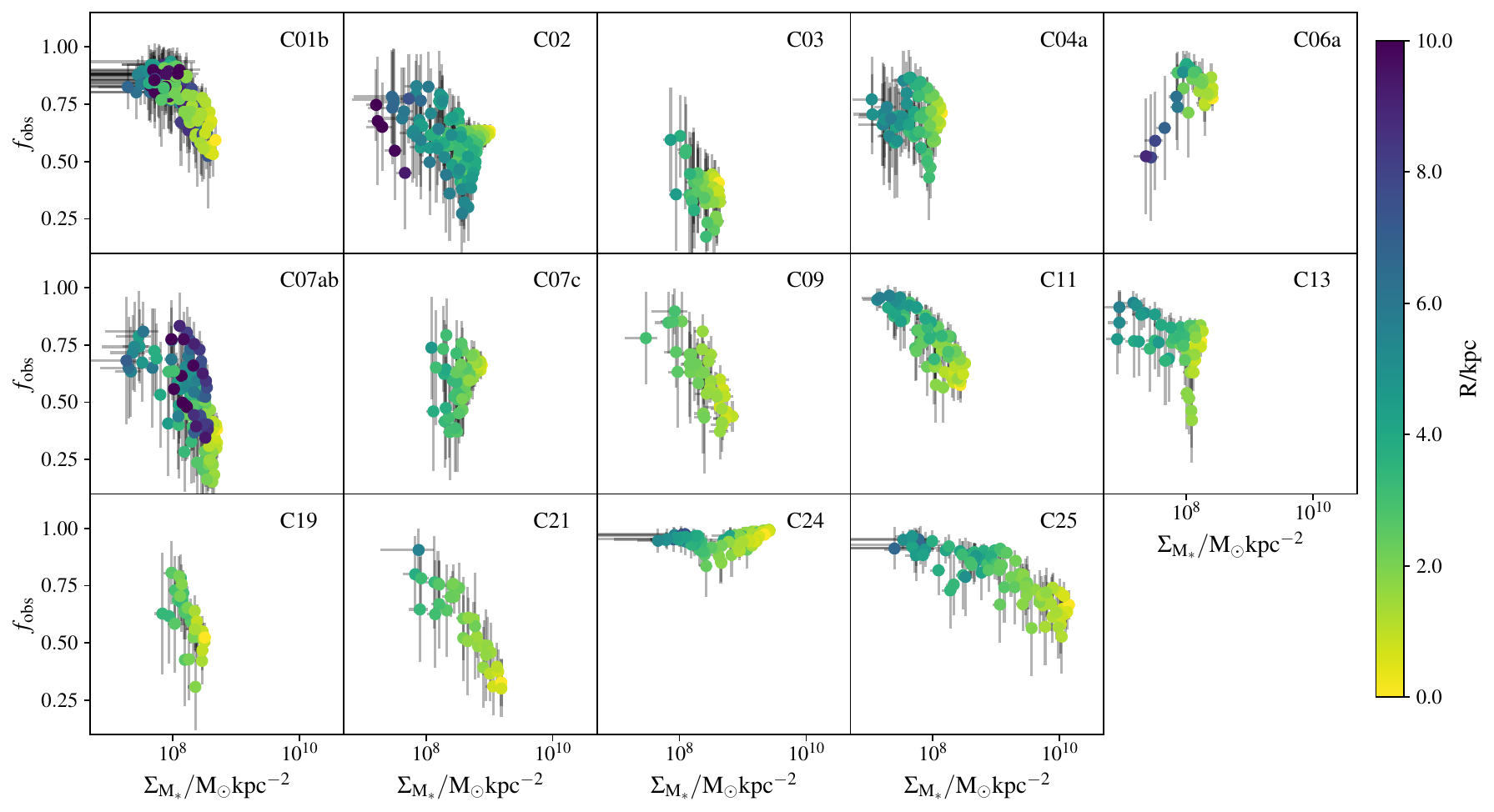}
    \caption{Fraction of dust-obscured SFR, $f_{\rm obs}$, plotted against the surface density of stellar mass, $\Sigma_{\rm M_*}$, for each $\sim0.5-1$ kpc spatial bin in our galaxies. The vertical errors are given by propagating the observational uncertainties in Eq.~\ref{eq:fobs}; the horizontal errors are given by \magphys. Each point is color-coded to reflect its distance to the brightest pixel in NIRCam F444W (which we use as an empirical proxy for the center of stellar mass).
    % CRISTAL-04b is not included in this analysis because these are too few spatial bins with detected ALMA flux to compute $f_{\rm obs}$. 
    }
    \label{fig:fobs-Mstar}
\end{figure*}

By studying their total, integrated emission, it has been found that galaxies with higher stellar masses tend to have higher dust obscuration\citep{GarnNBest2010,Battisti2022,Shapley2022}. 
A correlation between the total stellar mass of galaxies and the fraction of their star formation rate that is obscured, \fobs, has been found from the local Universe out to $z\sim7$ (e.g., \citealt{whitaker2017,Fudamoto2020a,Inami2022,Ferrara2022,Algera2023}), including for the CRISTAL sources \citep{Mitsuhashi2023}. 
Here we test if such a correlation exists at sub-galactic scales within our CRISTAL galaxies. To do so, we compare the surface density of stellar mass with the fraction of obscured star formation in our $\sim0.5-1$\,kpc spatial bins.

%We understand that most massive galaxies are enshrouded in dust, thus we should always take dust attenuation into account when studying SFR of massive galaxies. In fact, dust extinction introduces the largest systematic bias in estimating the global SFR of galaxies \citep{kennicutt1998a}. 
%To have a complete view of the total SFR, one way is to add the obscured SFR$_{\rm IR}$ estimated from mid-infrared (MIR) to far-infrared (FIR), with the unobscured SFR$_{\rm UV}$ measured at rest-frame UV continuum.
%Due to limited number of FIR and sub-mm selected galaxies to-date, \citet{whitaker2017} built an empirical relation between fraction of obscured star formation ($f_{\rm obs}$) and stellar mass which can then be used estimated the obscured star formation of galaxies at any given stellar masses.
%They found that $f_{\rm obs}$ is higher with higher stellar mass and lower at higher redshifts (up to z$\sim$2.5). 
%This positive correlation is also confirmed in the CRISTAL survey \citep{Mitsuhashi2023}, REBELS survey \citep{Algera2023} and ALPINE survey \citep{Fudamoto2020a} at redshifts 4 to 9.
%However, all of the above are integrated studies of the global $f_{\rm obs}$, or even stacking FIR of galaxies in the same stellar mass bin. We wish to unfold this relation in sub-galactic scales leveraging the high resolution data we have in both FIR and UV wavelengths.

Following previous works (e.g., \citealt{whitaker2017}), we define $f_{\rm obs}$ as:
\begin{equation}
    f_{\rm obs} = \frac{\rm SFR_{IR}}{\rm SFR_{IR}+SFR_{UV}} \,,
    \label{eq:fobs}
\end{equation}
where $\rm SFR_{IR}$ and $\rm SFR_{UV}$ are the star formation rates inferred from the infrared emission and from the ultraviolet emission, respectively. We compute $f_{\rm obs}$ empirically to avoid any circularity or potential degeneracy with our SED fitting. To compute $\rm SFR_{IR}$, we first convert the single-band ALMA dust continuum measurement (Band 7 corresponding to the rest-frame $\sim158\,\mu$m dust continuum) into a total IR luminosity using the stacked dust SED template of \citet{Bethermin2020} (i.e., at $158\mu m$, $\rm \nu L_{\nu}/L_{IR} = 0.133$), which is based on galaxies from  ALPINE (the parent sample of CRISTAL). Then, we convert the total IR luminosity to $\rm SFR_{IR}$ using the \cite{kennicutt1998a} conversion factor. To compute $\rm SFR_{UV}$, we use the rest-frame $\sim1800-2000\,$\AA\ UV flux measured from the JWST/NIRCam F115W images to obtain the observed (i.e., not dust-corrected) UV luminosity. Then, we convert the UV luminosity to $\rm SFR_{UV}$ using the conversion factor given in \citet{Madau2014} (for consistency, we divide both $\rm SFR_{IR}$ and $\rm SFR_{UV}$ by 1.6 because the SFR prescriptions are based on a \citealt{Salpeter1955} IMF, and \magphys\ adopts the \citealt{Chabrier2003} IMF).

%star formation rate inferred from IR flux (i.e. dust continuum emission from ALMA band 7) and UV flux (i.e. F115W NIRCam image) at restframe $\sim158\,\mu$m and $\sim1800-2000\,$\AA\ respectively. Thus, this quantity is SED modeling independent. 
%The conversion between dust continuum emission at 158$\,\mu$m to L$_{\rm IR}$ following \citet{Bethermin2020} which is computed based on stacking the FIR observations of all ALPINE galaxies (i.e. the parent sample of CRISTAL galaxies). Then, we converted L$_{\rm IR}$ to SFR$_{\rm IR}$, and L$_{\rm UV}$ SFR$_{\rm UV}$ following \citet{Madau2014}. We divided the obtained SFRs by a factor of 1.6 throughout this work to account for the difference between IMF models assumed in the \citet{Madau2014} (i.e. a \citet{Salpeter1955} IMF) and in \magphys (i.e. a \citet{Chabrier2003} IMF).
%We did not perform any extinction correction to the observed UV flux.\\

% Moreover, both agree that the dust emission seems to be more extended compare to the UV as the edge of the galaxy tend to be more reddened compare to the center where the stellar mass and star formation peaks.\\

In Fig.~\ref{fig:fobs-Mstar}, we plot \fobs\ of each spatial bin against the surface stellar mass density \sigmstar\ for all our CRISTAL sources.
Contrary to the known correlation for total, integrated quantities, we do not observe a clear positive correlation between \fobs\ and \sigmstar\ on resolved scales. In fact, they seem to be anti-correlated, with $f_{\rm obs}$ typically lower in the regions of highest \sigmstar.
This anti-correlation is also observed between \fobs\ and the surface density of SFR because of the strong correlation between \sigmstar\ and \sigsfr, as shown in Section~\ref{sec:rSFMS}. Fig.~\ref{fig:fobs-Mstar} also shows that \fobs\ tends to be higher further away from the center of the galaxy (defined empirically as the peak of the JWST/NIRCam F444W emission; we use this as a model-independent proxy for where the stellar mass peaks). This is somewhat at odds with findings of more centrally concentrated dust-obscured star formation found in other studies (e.g., \citealt{Fujimoto2017,Tadaki2020ApJ}), however, we note that \cite{Mitsuhashi2023} find typically more centrally concentrated UV emission than dust emission for CRISTAL galaxies.
We will further discuss these findings in Sect.~\ref{sec:fobs-r}.

%The coloring of the data points are based on the distance between the pixel to the peak in NIRCam F444W which is a proxy of where the stellar mass peaks.
%From the coloring of data points, we also noticed that $f_{\rm obs}$ tends to be higher further away from the center of the galaxy. This is again surprising as previous studies have found that the dust emission tend to be more compact than the optical/UV.
%We will further discuss our understanding of this finding in Sect.~\ref{sec:fobs-r}.
% This is also observed in the $f_{\rm obs}$ map and $\rm A_V$ map shown in Fig.~\ref{fig:maps} that the reddening is more sever towards the edge of the galaxy, though the morphology of CRISTAL-06a is more complicated.

\begin{figure}
    \centering
    \includegraphics[width=\linewidth]{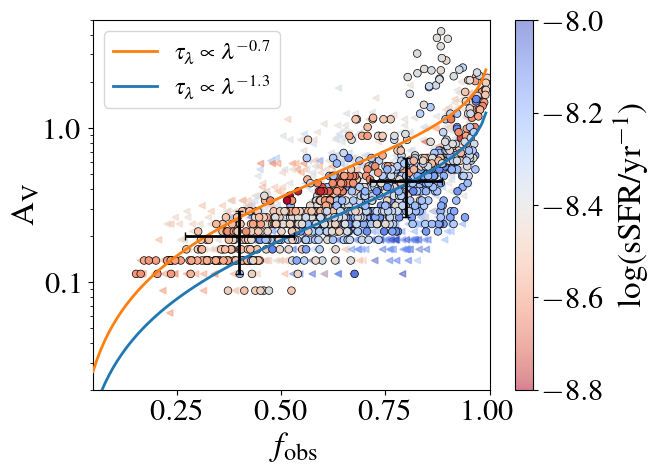}
    \caption{Average $V$-band dust attenuation given by \magphys, \av, plotted against the empirically-derived fraction of obscured star formation, \fobs, for all the spatial bins in our sample. The circles show regions where there were enough data to derive both quantities; the triangles show regions with upper limits on \fobs\ (i.e., points with only rest-frame UV but no ALMA detection). All regions are colored according to their specific SFR derived by \magphys. The orange and blue solid curves show the analytical relation predicted for pure diffuse ISM and birth cloud attenuation respectively. The two black error bars show the median errors for \fobs\ above and below 0.5, respectively.}
    \label{fig:fobs-Av}
\end{figure}

The fraction of dust-obscured star formation, \fobs, should be correlated to the average dust attenuation given, for example, by the $V$-band optical attenuation, \av. To check the consistency of our results, we compare the spatial distribution of the \magphys-derived \av\ with the empirically-derived \fobs. 
% We find that \av\ and \fobs\ both peak near the dust emission peak as traced by ALMA. 
We find that \av\ and \fobs\ tend to coincide with the dust emission spatially as expected. We can further compare them by checking if they are well correlated in our spatial bins. In Fig.~\ref{fig:fobs-Av}, we plot \av\ against \fobs\ for all the spatial bins in our sample where we had enough data to compute both quantities. Despite the relatively large uncertainties of individual data points, we find that \av\ and \fobs\ are well correlated, as expected.
We further derive an analytical relation between these two quantities:
\begin{equation}
    A_{\rm{V}} = -k\times \log_{10}(1-f_{\rm obs})\,,
    \label{eq:av}
\end{equation}
where $k$ depends on the power-law slope of the dust attenuation, and on the specific wavelength used to estimate the UV luminosity. For this calculation we adopt $\lambda_{\rm{UV}}=1900$\,\AA\, and we use the two-component dust attenuation modeled by \cite{Charlot2000} as given in \cite{magphys2008} (i.e., different dust attenuation slopes for the birth cloud and diffuse ISM components). Therefore, we find:
\begin{equation}
    k = \begin{cases}
    -1.19, &\, \text{if}\,\,\,\tau_{\lambda}\propto \lambda^{-0.7}\ \text{(diffuse ISM)}\\
    -0.63, &\, \text{if}\,\,\,\tau_{\lambda}\propto \lambda^{-1.3}\ \text{(birth cloud)}
    \end{cases}
    \label{eq:k}
\end{equation} 
%Here we adopted UV at $1900$\AA\ and V at $5500$\AA, and $\tau_{\lambda}$ is the optical depth at a given wavelength and it is related to the extinction parameter by $\rm A_{\lambda}=-2.5\times{\rm log_{10}}e^{-\tau_{\lambda}}=1.086\tau_{\lambda}$.
We plot this analytical prescription in Fig.~\ref{fig:fobs-Av}: it reproduces well the distribution and correlation shown by our CRISTAL data. We note that \magphys\ adopts an age-dependent dust attenuation as prescribed by \cite{Charlot2000}, wherein the light produced by young stars is attenuated by dust both in the birth clouds and in the diffuse ISM, and the light from older stars only suffers attenuation from diffuse ISM dust. Therefore, it is, in practice, a combination of both dust curves in an age-dependent manner. One would expect the effective dust attenuation curve of galaxies dominated by young stars to be closer to the birth cloud curve, while that of galaxies dominated by older stars to be closer to the diffuse ISM curve. As Fig.~\ref{fig:fobs-Av} shows, we find indeed that, overall, regions in our galaxies with higher specific SFRs (i.e., younger stellar ages) are closer to the birth cloud attenuation curve in the \fobs--\av\ plane, while regions with lower sSFRs (i.e., older stellar ages), are closer to the diffuse ISM component. The scatter seems to be dominated by the relatively large measurement uncertainties.

\section{Discussion}\label{sec:discussion}

\subsection{Resolved vs. integrated properties}

\begin{figure}
    \centering
    \includegraphics[width=\linewidth]{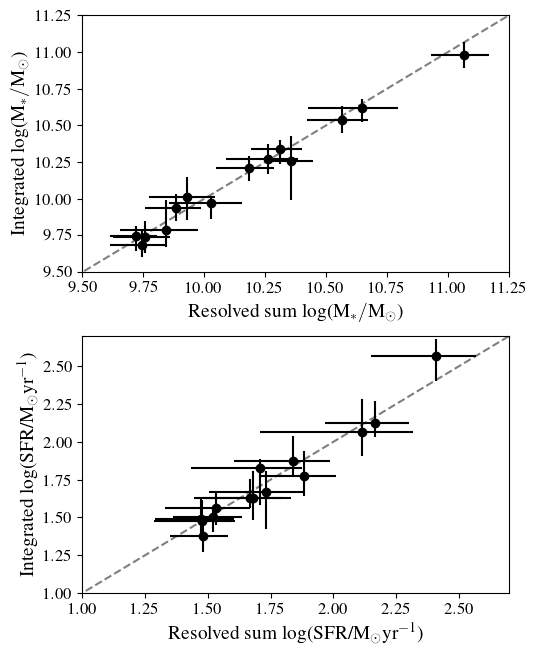}
    \caption{
    Comparison between the resolved sum and integrated stellar mass (top panel) and SFR (bottom panel) of the CRISTAL galaxies. There are no offsets between the integrated and resolved properties.
    }
    \label{fig:resolved-integrated}
\end{figure}

Previous spatially-resolved studies have found that integrated SED fitting may underestimate the stellar masses of galaxies (e.g., \citealt{Sorba2015,Sorba2018}). In this section, we explore how our resolved approach compares to the results derived from the traditional integrated methods.
% We plotted the SFR against stellar mass of the CRISTAL galaxies inferred from two different methods in Fig.~\ref{fig:SFMS}. In red, we compute the total SFR and sellar masses of each galaxies by summing up the values in each independently modeled pixels, with error being the median fractional error of all pixels (i.e. resolved). In yellow, we first sum up the SEDs of all pixels, then modeled the total SED with \magphys to obtain the physical parameters and uncertainties (i.e. integrated). The integrated approach is essentially equivalent to the traditional aperture photometry with a fixed (total) aperture across all filters.
% In the main panel, with either method we can see the CRISTAL galaxies fall on the star forming main sequence of $4<z<6$ as represent by the grey crosses taken from the ALPINE survey (see \citealt{Bethermin2020,Faisst2020}, etc.), which is the parent sample of the ALMA CRISTAL survey.
% In the two side panels, we compared the inferred SFRs and stellar masses with each methods of each CRISTAL galaxy by subtracting the `integrated' from the `resolved'. We found that the difference is often no more than 0.1 dex which is consistent with no difference given the error bars.
% We also analysed the results of these two methods by fitting the slope of star forming main sequence (i.e. SFR$\propto \rm M_*^\alpha$) inferred by each method which are shown by the dashed lines. We found that the slope for resolved approach is shallower but is well within uncertainty.

In Fig.~\ref{fig:resolved-integrated}, we compare the sum of the stellar masses and SFRs in all of our individual spatial bins for each galaxy, with the values derived by fitting the total, integrated fluxes with \magphys. Both the stellar masses and SFRs agree remarkably well between the two methods, showing that, at least for this dataset and SED model, the integrated analysis does not suffer from biases, as found in previous studies. We note that we also do not find any systematic differences when comparing to the properties derived by \cite{Mitsuhashi2023} using a different SED model (i.e., CIGALE; \citealt{Boquien2019}) and different sets of input integrated photometry. We do not observe the discrepancy in stellar mass reported in \citet{Sorba2018}, who found that integrated SED fitting can underestimate galaxy stellar masses up to a factor of five, with larger offsets for higher specific SFR. They attributed this to mass-to-light ratios being underestimated in their global fitting due to outshining of young stellar populations (see also, \citealt{Maraston2010}).  
The resolution of our observations, set by the ALMA beam ($\lesssim 3$\,kpc), is at the limit where \citet{Sorba2018} argue one would no longer find a discrepancy. However we note that, in a recent study, Lines et al. (in prep.) also find no outshining in a sample of four CRISTAL galaxies where spatially-resolved SED fitting was performed at higher resolution using using a different SED model but not including ALMA observations.
% The reason our method does not suffer from this bias may be the wide wavelength coverage of our observations (including ALMA, which helps break the age-dust degeneracy).

%%% MOVE?
% We also infer the SFR and stellar mass with both resolved and integrated method with only HST and JWST filters which is the most common situation in SED analysis of high redshift galaxies (e.g., \citealt{Cibinel2015, Giménez-Arteaga2023,Song2023}). The results of resolved and integrated analysis are again consistent with each other without ALMA observations.
% Nonetheless, later we will show the inferred SFR and stellar mass with and without ALMA observation, no matter resolved or not, may not be always consistent.

% Nonetheless, we found that the fitted slope is now $\sim1$ which is higher than with ALMA for both resolved and integrated approaches. 
% This is a result of over estimation of SFR for high stellar mass CRISTAL galaxies ($\rm log(M_*)\gtrsim 10 M_{\odot}$) by $\sim0.2$ dex. 
% We note that the over estimation of SFR are still consistent with the uncertainty but systematically higher with higher stellar mass.
There are often concerns that the energy balance assumed by some SED modeling codes, including \magphys, may no longer be valid when spatial UV/IR offsets exist. We find that the `resolved' and `integrated' approaches using the same version of \magphys\ return similar global values for all the CRISTAL sources, even the ones with spatial offsets between the UV and IR emission. We can conclude that the energy balancing assumption adopted in \magphys\ still works regardless whether there is an UV-FIR spatial offset; this could be attributed to \magphys\ having some degree of flexibility in the energy balance criterion (the energy re-radiated by dust has to be within 15\% of the energy absorbed; \citealt{magphys2008}). This finding is consistent with a recent independent study by \citet{Haskell2023} who tested \magphys\ on using high-resolution simulated galaxies with spatially-decoupled UV and FIR emissions, and found that the fidelity of \magphys\ is independent of the degree of UV/FIR offset provided that quality of fit is acceptable.

%Nevertheless, resolved studies would naturally mitigate the concerns of integrating and balancing physically decoupled UV and FIR emissions, since, on smaller scales, if UV and FIR emission are both present, they are more likely coupled to each other.

\subsection{The impact of ALMA observations}

\begin{figure*}
    \centering
    \includegraphics[width=0.9\linewidth]{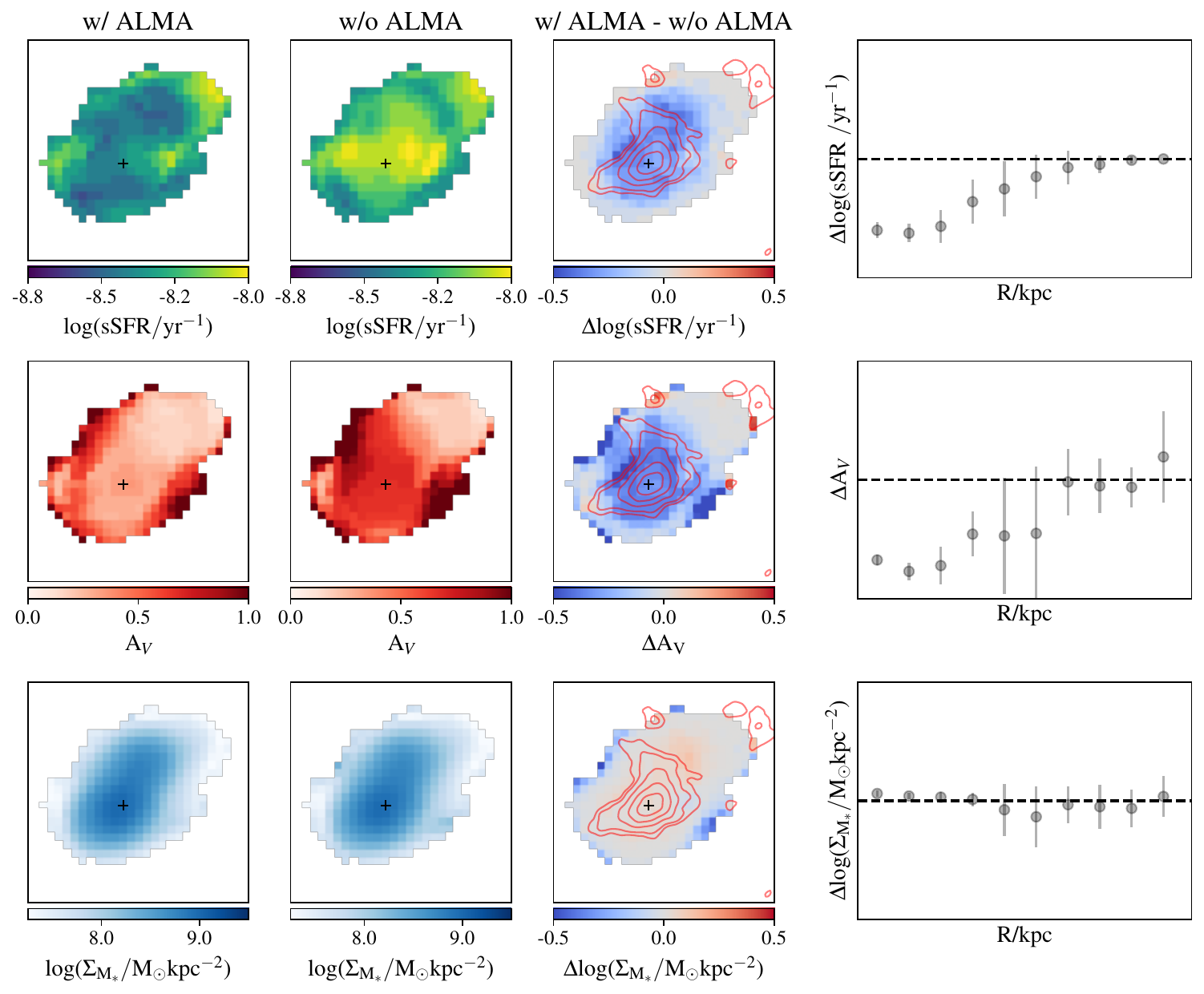}
    \caption{Comparison of the results of our spatially-resolved SED fitting for CRISTAL-02 when ALMA observations are included (1st column) and excluded (2nd column). The 3rd column show the deviation by subtracting the 1st by the 2nd column, with red contours showing the dust emission at 2-, 3-, 5-, 7- and 9-$\sigma$. The last column show the radial distribution of the median deviation per radial bin measured from the center of galaxy (marked as a black cross). The three rows show, from top to bottom, specific SFR (sSFR), $V$-band dust attenuation (\av), and surface stellar mass density ($\Sigma_{\rm M_*}$).}
    \label{fig:compare-no-ALMA}
\end{figure*}

\begin{figure}
    \centering
    \includegraphics[width=1\linewidth]{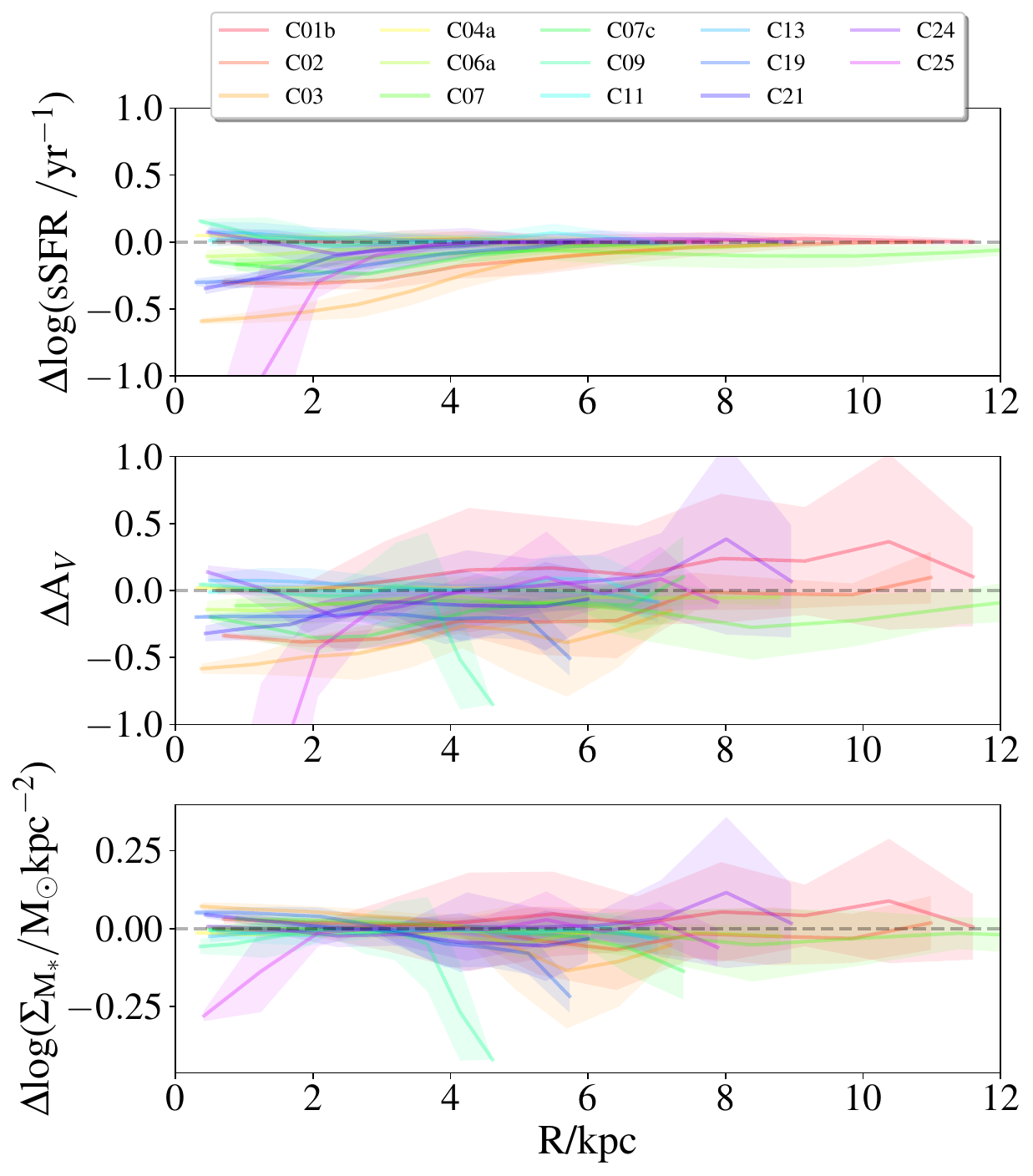}
    \caption{Radial distributions of the median deviation in sSFR, \av, and \sigmstar\ per radial bin measured from the center for all 14 sources, comparing the spatially-resolved SED fits with and without ALMA (similar to the right-hand column of Fig.~\ref{fig:compare-no-ALMA}). The shaded regions indicate the standard deviation for each radial bin.}
    \label{fig:all-wwo-ALMA}
\end{figure}

Given the unprecedented image resolution and depth afforded by combining HST and JWST, recent studies have performed spatially resolved SED fitting on high-redshift galaxies. However, most such studies lack information about the FIR dust emission on similar scales (e.g., \citealt{Abdurrouf2022,GimenezArteaga2023,Kokorev2023,Song2023}; though see \citealt{Smail2023}, who also include sub-millimetre constraints albeit at poorer resolution). Here, we explore the influence of the spatially-matched ALMA observations on our SED fitting.
In Fig.~\ref{fig:compare-no-ALMA}, we compare, for source CRISTAL-02, the results of our spatially-resolved SED fitting with and without including the ALMA dust continuum observations. In both cases, all other settings remain unchanged (e.g., resolution, binning, \magphys\ version). We show the spatial distribution of three physical parameters (specific SFR, \av, and \sigmstar) and the radial distributions of the differences between the two cases.
For CRISTAL-02, including ALMA data changes the inferred specific SFR map dramatically. The sSFR is higher almost everywhere when not including ALMA observations, while with ALMA observations the sSFR is overall lower with a few local maxima. In this case, without ALMA, the radial distribution of sSFR is more centrally concentrated, while with ALMA it is flatter. These differences could be important when assessing `inside-out' versus 'outside-in' galaxy assembly scenarios (e.g., \citealt{Tacchella2015,Tacchella2018,Nelson2019}). We note that both distributions show a slight elevation at higher radius, which reflects the blueish feature north-west of the system with no significant ALMA emission detected (Fig~\ref{fig:maps}).

To better understand these differences in sSFR, we compare the dust attenuation (\av) maps obtained with and without ALMA observations.
The changes in \av\ between the two cases are more subtle, but Fig.~\ref{fig:compare-no-ALMA} shows that \av\ is enhanced in the central region of the galaxy when ALMA data are not included. 
% This hints that the difference between sSFR is due to the age-dust degeneracy. 
\textcolor{black}{In the case of no ALMA observations, relatively red HST and JWST colors are interpreted by the model as having younger ages, higher \av, \ldust\ and dust temperature ($T_{\rm dust}$); when ALMA observations are included, they help constrain $T_{\rm dust}$, \ldust\ and \av\ via the energy balance, and hence the model prefers slightly older ages, and lower sSFR. }

\textcolor{black}{We also investigate the impact of ALMA observations on the inferred stellar mass surface density, \sigmstar. We find that, when ALMA data are not included in the fits, because of the bias toward lower galaxy age, the stellar masses are also slightly lower compared to the case where ALMA data are included. However, the difference in stellar mass is negligible because stellar mass is mostly constrained by the rest-frame optical JWST/NIRCam observations. When combining the effect of similar stellar masses and higher in the sSFR, the total SFR is higher for CRISTAL-02 by roughly $0.2$\,dex (from $\sim135\rm~M_{\odot}yr^{-1}$ with ALMA to $\sim230\rm~M_{\odot}yr^{-1}$ without ALMA).}

We repeated this test for all the galaxies in our sample. In Fig.~\ref{fig:all-wwo-ALMA}, we show the radial distributions of the median deviation in sSFR, \av, and \sigmstar\ per radial bin measured from the center for all 14 sources. Overall, we find that not including ALMA changes the radial distribution of specific SFR for half of our sample. \textcolor{black}{In these cases, the central sSFR tends to be higher when ALMA observations are not included, as described for CRISTAL-02. This can lead to difference of the central sSFR by up to a factor of three. This is related to systematic offsets in the inferred \av\ as shown in the middle panel of Fig.~\ref{fig:all-wwo-ALMA} essentially, because ALMA helps constraint the \ldust\ and $T_{\rm dust}$, as discussed above.} The stellar masses are relatively robust because they are mostly constrained by the optical/near-IR observations.

\begin{figure*}
    \centering
    \includegraphics[width=0.9\linewidth]{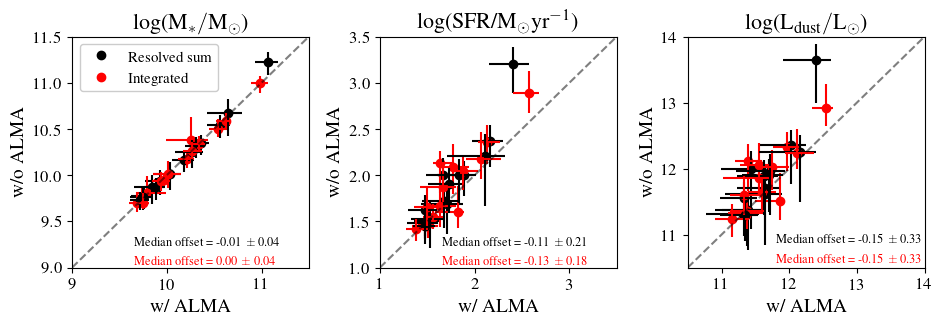}
    \caption{Comparison of physical parameters obtained with and without including ALMA observations in the SED fitting. From left to right, the panels show the comparisons of stellar masses, star formation rates, and dust luminosities. In each panel, the black points compare the results from the resolved sum approach (i.e., adding up the parameters in all the spatial bins) and the red points compare the results from the integrated approach (i.e., fitting the total integrated SED).}
    \label{fig:with_without_alma}
\end{figure*}

To test the effect of ALMA observations on our whole sample, in Fig.~\ref{fig:with_without_alma}, we compare the inferred values for stellar mass, SFR and dust luminosity, using both the resolved sum and integrated approaches, with and without ALMA. The stellar masses inferred with and without ALMA observations are overall consistent with each other for both resolved and integrated approaches. This is likely because the stellar mass is mostly constrained by the JWST filters at rest-frame optical/NIR wavelengths. The SFR and dust luminosity are, however, not fully consistent when ALMA observations are excluded from the SED modeling.
\textcolor{black}{Excluding ALMA leads to a higher estimation of the \ldust\ (by about 0.15 dex) with a slight bias towards higher SFR, although the values are still broadly consistent given the larger error bars. The estimated SFR can be also higher, most likely a result of larger inferred dust-obscured star formation rates resulting from the higher \ldust. We found that whether there are deviations on estimated SFR and \ldust\ depends on whether the inferred $T_{\rm dust}$ changes with the additional ALMA observation. There are cases that even having the ALMA observation, the inferred $T_{\rm dust}$ still closely resembles the centre of prior of $T_{\rm dust}$. In these cases, it seems that a single ALMA observation is not sufficient to provide constraint on $T_{\rm dust}$, \ldust\ and \av\ such that the inferred age and SFR are consistent with or without ALMA.}
\textcolor{black}{Our findings are consistent with \citet{Pacifici2023} (e.g. their Figure 5) who also found the results when not including FIR observations are slightly biased to higher SFR and \av\ for \magphys\ as well as 4 other different SED fitting codes over a larger sample of galaxies.}
We note that the galaxy with the highest SFR and \ldust\ is the strongest outlier in resolved approach. This is CRISTAL-25, for which we find that the SED may be contaminated by AGN line emission in the central region; we discuss this source in detail in Appendix~\ref{app:C25}.

\textcolor{black}{Including ALMA also noticeably improved the precision on estimating SFR, \av, \ldust\ and sSFR for over half of the galaxy in our sample. However, galaxy with more HST and JWST filters (i.e. $>10$) in the SED tend to have smaller or negligible differences. We think the reason could be HST and JWST filters dominate the $\chi^2_{\mu}$ such that a single ALMA observation does not have a large improvements on the precision. Having a high sampling of rest-frame far UV filters also helps in breaking the age-dust degeneracy to some extent because of the ability to constrain the shape of the required dust attenuation curve (e.g., \citealt{Pope2023}, see their Figure 2).}

\subsection{What could be causing the radial variation of \fobs?}\label{sec:fobs-r}

\begin{figure*}
    \centering
    \includegraphics[width=1\linewidth]{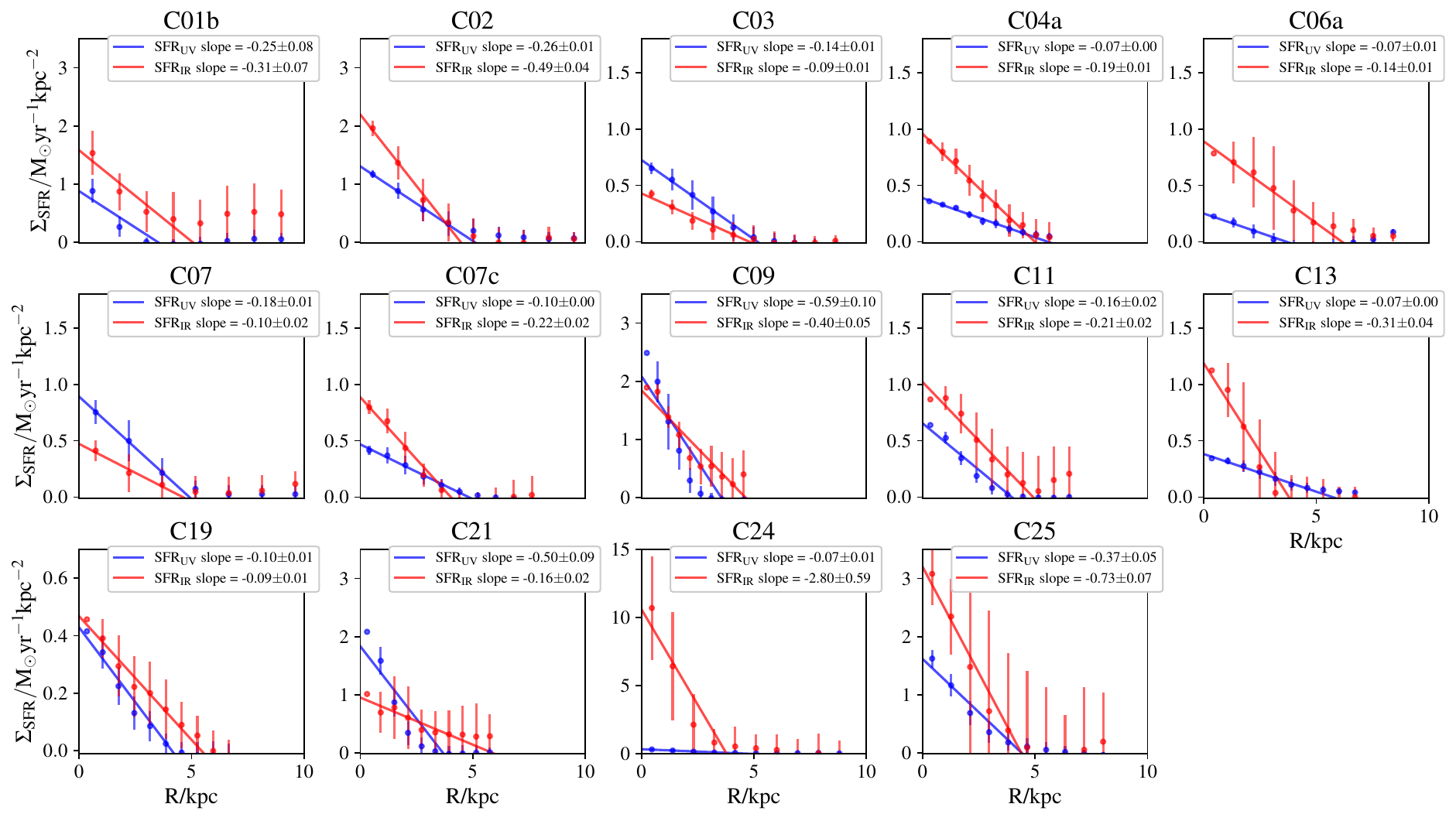}
    \caption{Radial distribution of the surface density of $\rm {SFR_{UV}}$ (blue) and $\rm {SFR_{IR}}$ (red) for all our CRISTAL galaxies. The error bars reflect the scatter within each radial bin. Straight lines show the best linear fits to the radial distributions. The best-fit values for the slopes are shown in the legend. }
    \label{fig:SFR-radial}
\end{figure*}

Our findings in Section~\ref{sec:fobs-trend} indicate a potential positive correlation between the fraction of dust-obscured star formation, \fobs, and radial distance away from the peak of the stellar mass in our galaxies.
% Two observations:
% (1) Dust missing from the center 
% -> (pressure/radiation?) feedback from star formation (The fobs at high radius may be at its normal value given the stellar mass)
%  Pro: Where would the dust go? Hard to not produce dust for the stellar mass, and SF still going on, why low in FIR at center where the SNR are not bad for both UV and FIR?
%  Con: We don't see significant outflow in [CII] line kinematic, though [CII] is at leaast as extended as the dust continuum.
% (2) Extended FIR emission at high radius 
% -> interaction causing dust-stars misalignment (explained the complicated radial trend and fobs is not higher isotropically at higher radius).
% -> not related to in situ star formation (clumpy ISM, high f_esc, UV photons escape to the IGM from the stellar disk. Heating up dust (and/or excited [CII]) in the IGM -> reionization related)
% -> energy balancing still work?
% The low overall fobs in general and low resolved fobs at center may be a results of selection effect. Higher fobs at outskirts are at boundaries between ISM and IGM.
To better understand this trend, we investigate the radial distribution of the surface densities of UV ($\rm \Sigma_{SFR_{UV}}$) and infrared ($\rm \Sigma_{SFR_{IR}}$) star formation rates in Fig.~\ref{fig:SFR-radial}, as these are the two empirically-derived quantities used to compute \fobs. Fig.~\ref{fig:SFR-radial} shows that $\rm \Sigma_{SFR_{UV}}$ and $\rm \Sigma_{SFR_{IR}}$ always peak at the center, and then radially decrease, though this radial decrease shows a large variation between our sources. There is particularly large scatter in $\rm \Sigma_{SFR_{IR}}$ in some sources, which may be caused by a combination of lower signal-to-noise ratio of the observations, as well as offsets between stellar and dust emission, and multiple peaks/clumps in the IR emission.
We linearly fit the radial gradients of $\rm \Sigma_{SFR_{UV}}$ and $\rm \Sigma_{SFR_{IR}}$ within 5\,kpc and find that, for about a third of the sample, $\rm \Sigma_{SFR_{UV}}$ shows a steeper slope than $\rm \Sigma_{SFR_{IR}}$, i.e., the UV SFR decreases faster with radius. 
% For the other galaxies with steeper slope for $\Sigma_{SFR_{IR}}$, one can observe higher $\Sigma_{SFR_{IR}}$
This could explain why \fobs\ tends to increase with radius for some of our sources. 

\citet{Mitsuhashi2023} found similar results when comparing the physical sizes of CRISTAL galaxies as traced by the UV and dust continuum emission. They found that the dust emission is comparable to or slightly more extended than the UV emission, for both individually detected galaxies and a stack of all undetected galaxies. 
Here we find that the dust emission is not simply isotropically more extended than the UV emission, but sometimes appears to extend to high radius in a particular direction. Dust emission at high radius where the UV emission is low would result in higher \av\ and \fobs\ (see Fig.~\ref{fig:maps} for an example).

One way to explain the extended dust emission is there is a `dust deficit' in the central regions of some of the CRISTAL galaxies, such that the \fobs\ are not as high as at outer regions for a given \sigmstar.
We speculate that the reason could be dust being removed from the center of these galaxies by star formation (massive stars and/or supernovae) feedback.
This picture is plausible as the central region is where \sigmstar\ and \sigsfr\ peak, and both simulations and low-redshift resolved studies indicate that feedback should correlate with \sigsfr\ (e.g., \citealt{Hopkins2012,ReichardtChu2022,ReichardtChu2024}).
In particular, the simulations of \citet{Arata2019} show that, at high redshifts ($z>6$), stellar feedback can completely remove the ISM to let UV photons escape free, and the peak of the SED rapidly changes between UV and IR on time scales $\sim100$\,Myr due to intermittent star formation and feedback. 
\cite{Ferrara2023,Ferrara2024} also proposed a model where dust can be efficiently ejected by outflows at $z>10$, which successfully explains the excess of blue bright JWST galaxies.
As we select massive star forming galaxies that are UV-bright ($\rm -22.5<M_{UV}<-21$) and also detected in the dust continuum by ALMA, we may be preferentially selecting galaxies in a phase that occurs soon after a starburst-driven outflow that removes dust from the central regions of the galaxy (e.g., \citealt{Lapi2018a}), allowing the detection in both UV and dust emission simultaneously.
A caveat to this scenario is that we do not observe significant signs of outflows in the [CII] line profiles of the majority of the CRISTAL galaxies, even though stacking [CII] emission of ALPINE galaxies supports outflows \citep{Ginolfi2020a}. It could be that the [CII] line is not an ideal tracer of molecular outflows (e.g., \citealt{Levy2023}). Observations of alternative outflow tracers (e.g., OH $119\mu$m; \citealt{Salak2024}) would be needed to verify this.

Another possible way to redistribute dust/gas from galaxies is through mergers or interactions.
Many CRISTAL galaxies consist of multiple clumps visually identified in either UV \citep{Mitsuhashi2023} and/or [CII] emission \citep{Ikeda2024}, and nearly half are classified as merging systems through spatially-resolved [CII] kinematics (Lee et al., in prep.). 
We find that \fobs\ at large radii tend to be higher where $\Sigma_{\rm sSFR}$ is higher (e.g., see Fig.~\ref{fig:fobs-Av}). Obscured star formation at the outskirts could be associated with dusty starbursts triggered by mergers, interactions or accretion. However, whether a galaxy consists of multiple clumps and/or is classified as a merger based on [CII] line kinematics does not predict if the resolved \fobs\ has a negative gradient. For example, both CRISTAL-24 and -25 have no clumps in UV, optical or [CII] images, but are classified as merging based on kinematics, and they show different \fobs\ to \sigmstar\ relations in Fig.~\ref{fig:fobs-Mstar}.

It is important here to explore a caveat of our measured anti-correlation between \fobs\ and \sigmstar: the dust temperature (and hence the dust SED) could vary spatially, which might invalidate our use of the constant flux-to-luminosity conversion factor from \citet{Bethermin2020} to obtain the total IR luminosity (see also the discussion in \citealt{Mitsuhashi2023}). If the dust temperature has a negative radial gradient, the inferred dust luminosity may be underestimated in the center and overestimated in the outskirts. This could lead to a less pronounced radial variation of $\rm \Sigma_{SFR_{IR}}$, and hence of \fobs. To quantify this, we simulated how a dust temperature gradient would affect the inferred \fobs\ distribution of CRISTAL-02 in Appendix~\ref{app:Tdust}. We find that even a steep temperature gradient between 65\,K in the center and 20\,K in the outskirts does not change the negative trend between \fobs\ and \sigmstar\ in that galaxy, with only the central region deviating significantly from this trend. Nevertheless, this could explain the complexity and variability among the sample, because dust temperature gradients might also vary from galaxy to galaxy. Multi-frequency matched-resolution ALMA observations (especially including higher frequencies; e.g., \citealt{Villanueva2024b}) would help constrain the dust temperatures more robustly (e.g., \citealt{Hodge2020}) and hence are key to obtaining better constraints on spatially-resolved dust obscuration in galaxies, and elucidate the relations between stellar content, star formation activity and dust obscuration on kpc-scales hinted by our analysis.

\subsection{Comparison with empirical SFR indicators}

\subsubsection{UV+IR star formation rates}

\begin{figure*}
    \centering
    \includegraphics[width=1\linewidth]{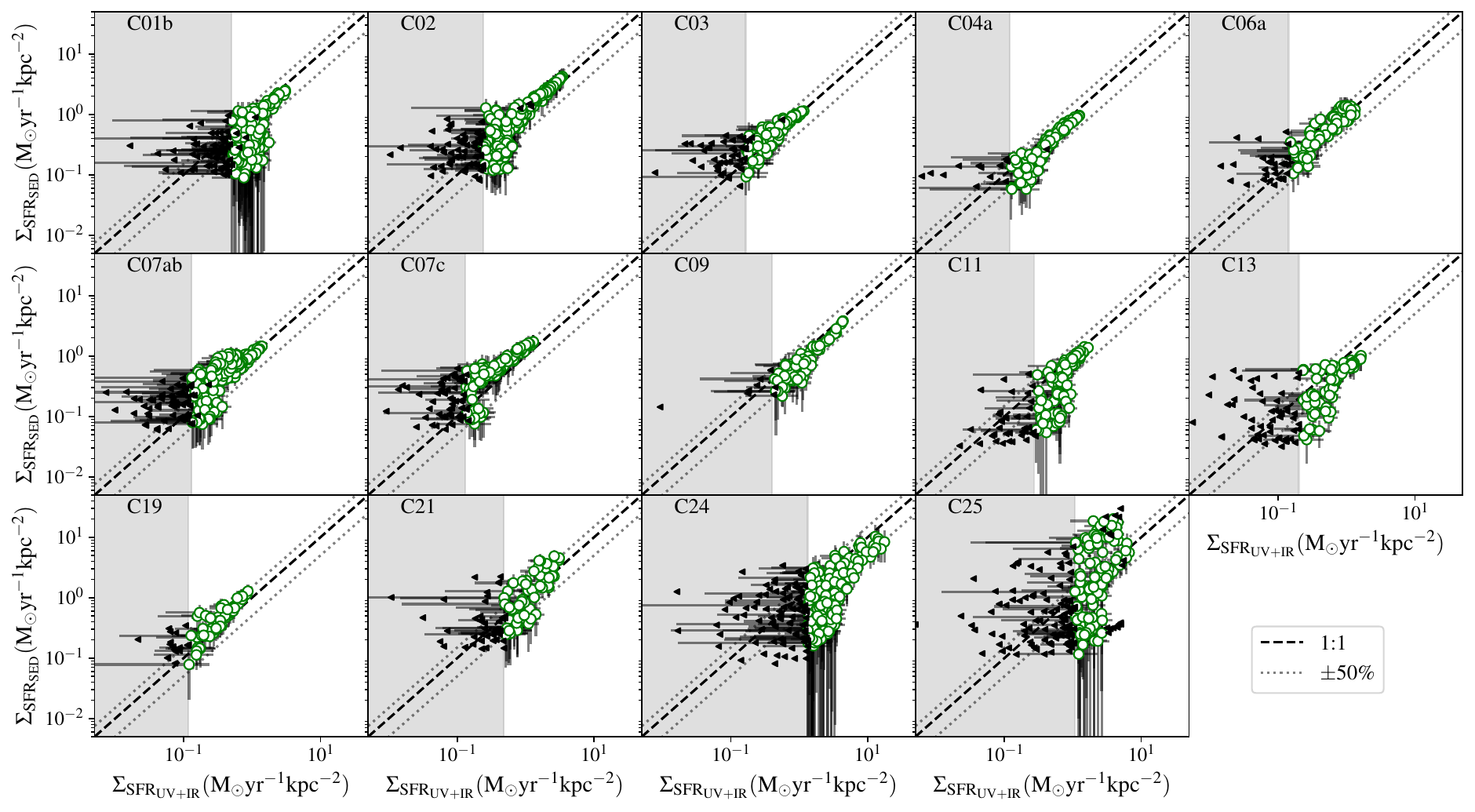}
    \caption{Comparison between the star formation rate surface densities obtained using \magphys\ (${\rm SFR_{SED}}$) and using UV+IR empirical indicators (${\rm SFR_{UV+IR}}$), for each spatial bin in our analysis. The green circles show regions where both ${\rm SFR_{SED}}$ and ${\rm SFR_{UV+IR}}$ could be reliably constrained (i.e., good SED fits and detections both in the UV and IR), and the black triangles are regions that do not fully meet the detection criteria. The grey area in each panel shows the $1\sigma$ detection limit of $\Sigma_{\rm SFR_{UV+IR}}$. The black dashed line in each panel shows the 1-to-1 relation, with dotted grey lines marking the $\pm50\%$ range.}
    \label{fig:SFR-SFR}
\end{figure*}

% The fixed IR flux to IR luminosity conversion factor assumed in this work is not ideal but inevitable. 
We now compare our spatially-resolved SED-based star formation rate maps with those derived using empirical SFR indicators based on the UV+IR emission and on the [CII] emission.

In Fig.~\ref{fig:SFR-SFR}, we first compare, for each spatial bin in each galaxy, the SFR inferred from our \magphys\ SED modeling to the total star formation rate $\rm SFR_{UV+IR}=SFR_{UV}+SFR_{IR}$, where $\rm SFR_{UV}$ and $\rm SFR_{IR}$ are computed as described in Section~\ref{sec:fobs-trend}. We find that the two SFRs are well correlated and close to the identity line for all galaxies, with a maximum offset of $\pm 50\%$ (except for CRISTAL-25, which along with having a strong UV/IR spatial offset, also may be contaminated by AGN flux in the center where SFR is at maximum; see Appendix~\ref{app:C25}). The scatter around the unity line is larger in regions where the SFR is lower, i.e., in the outskirts of galaxies, where the surface brightnesses are typically lower. This scatter is likely not physical, but rather due to the low SNR of the observations and larger resulting errors on SFR.

% In CRISTAL-04a and -11, $\rm SFR_{UV+IR}$ tend to be higher than $\rm SFR_{SED}$ when $\rm \Sigma_{SFR_{SED}}\lesssim0.1M_{\odot}yr^{-1}kpc^{-2}$. When we take the upper limits into account,  $\rm SFR_{SED}$ tend to be higher than $\rm SFR_{UV+IR}$ in CRISTAL-03, -06a and -13 when $\rm \Sigma_{SFR_{UV+IR}}\lesssim0.1M_{\odot}yr^{-1}kpc^{-2}$. 

%the surface SFR density inferred from two single band SFR indicators (i.e. $\rm SFR_{UV+IR}$) on the x-axes and the surface SFR density from SED fitting on the y-axes (i.e. $\rm SFR_{SED}$). The black dashed line shows the 1-to-1 ratio. The grey area in each panels marked the regions of the 1$\sigma$ limit of our estimation of SFR$_{\rm UV+IR}$. Data point fall below this limit are not statistically detected. Also, we only plot resolved element that are at least 1$\sigma$ detected in UV and IR filters and acceptable SED fitting results ($\chi^2<2$) as green circled data points, otherwise they are represented by black triangles.
%A $\sim\pm50\%$ range is indicated by the two faint grey lines above and below the black dashed lines in Fig.~\ref{fig:SFR-SFR}. 

It is worth noting that the empirical SFRs may suffer from systematics. As we use the same filter (JWST/NIRCam F115W) to compute as $\rm SFR_{UV}$ across the whole sample, there could be small systematic offsets caused by the variation in rest-UV wavelength probed at different redshifts (i.e., from $\lesssim1800$\,\AA\ to $\simeq2000$\,\AA).
Additionally, $\rm SFR_{IR}$ may suffer systematic uncertainties from the use of a single-band ALMA measurement to infer the total IR luminosities, as discussed in the previous section.
The SED-based SFRs are less likely to be affected by such systematics, because they take into account the full observed SED, and also because they implicitly assume a much wider range to UV-to-SFR and IR-to-SFR thanks to the wide priors of star formation histories and dust properties used in \magphys.

\subsubsection{[CII] star formation rates}

\begin{figure*}
    \centering
    \includegraphics[width=\linewidth]{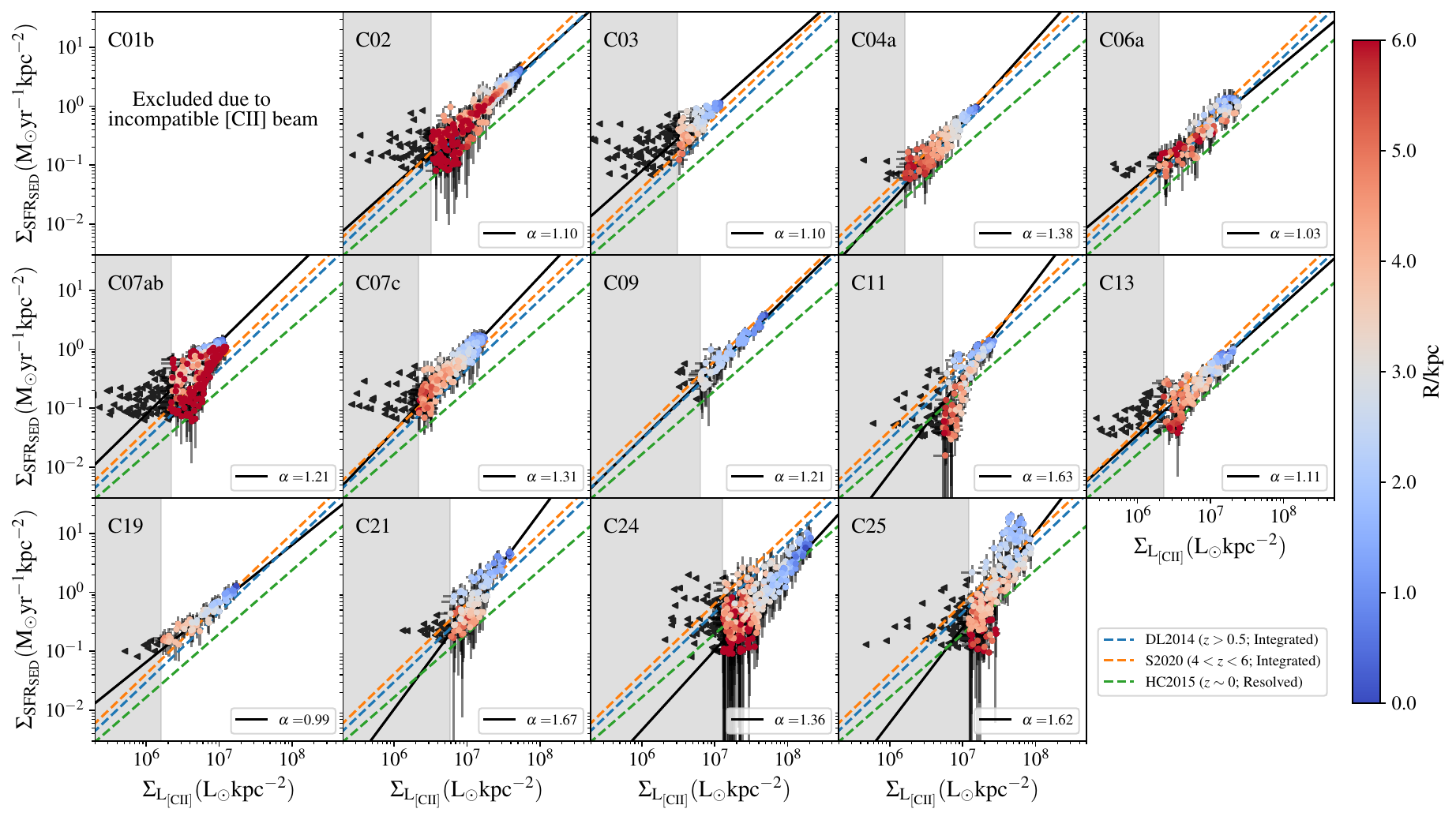}
    \caption{Resolved SFR--$\rm L_{[CII]}$ relations. We compare the surface density of SFR with the surface density of [CII] luminosity in each of our spatial bins that have $3\sigma$[CII] detection and reliable SED fits. Each point corresponds to a spatial bin and is colored according to distance to the center (i.e., the peak of JWST/NIRCam F444W emission). The grey area shows the $3\sigma$ detection limits of $\Sigma_{\rm L_{[CII]}}$. The black triangles are spatial bins that fall below the detection limit of [CII], but for which can still measure SFR. The solid black lines show the best-fit power law to only data points at $R<4$\,kpc, with the slope $\alpha$ indicated in legend of each panel. The colored dashed lines are the SFR--$L_{\rm [CII]}$ relations obtained from integrated measurements of intermediate- and high-redshift galaxies (DL2014 and S2020, \citealt{DeLooze2014,Schaerer2020}) and a local resolved study (HC2015, \citealt{Herrera-Camus2015}). We exclude CRISTAL01b from this analysis because beam size of the [CII] observations is larger than the beam size of dust continuum map used in the spatially-resolved SED fitting.}
    \label{fig:LCII-SFR}
\end{figure*}

The [CII]158$\mu$m emission line is the main coolant in neutral gas and traces the heating on small dust grains by UV photons \citep{Hollenbach1999}. It has been proposed as a SFR indicator in galaxies, especially at high redshifts ($z>2$), where it is readily detected by ALMA. To this end, several calibrations of SFR based on the total [CII] luminosity have been obtained, both on integrated and resolved scales (e.g., \citealt{delooze2011, Herrera-Camus2015,DeLooze2014, Herrera-Camus2018ApJ,Schaerer2020}). 
Since [CII] has been targeted for all our CRISTAL galaxies with ALMA, we test how well the [CII] luminosity traces our SED-derived SFRs on $\sim$kpc scales at $z>4$.

We compute the total [CII] luminosity $L_{\rm [CII]}$ from the observed flux density $S_{[\rm CII]}$ following \citet{Solomon2005}:
\begin{equation}
    L_{\rm [CII]} /  L_{\odot} = 1.04 \times 10^{-3} S_{\rm [CII]} \Delta v D_{L}^2\nu_{\rm rest}(1+z)^{-1},
\end{equation}
where $S_{\rm [CII]}$ is the flux density in Jy, $\Delta v$ is the line width in km\,s$^{-1}$, $D_L$ is the luminosity distance in Mpc, and $\nu_{\rm rest}=1900.54$\,GHz.

In Fig.~\ref{fig:LCII-SFR}, we compare, for each spatial bin, the surface density of [CII] line luminosity, \sigcii, with the surface density of SFR derived using \magphys, $\rm \Sigma_{SFR_{SED}}$, and find that they correlate well with each other at the $\lesssim$\,kpc scales probed here, albeit with significant scatter. The worst correlations are found in CRISTAL-24 and CRISTAL-25, which have spatial offsets between the peaks of \sigsfr\ and [CII] line luminosity. Note that the central region ($R\lesssim2$\,kpc) of CRISTAL-25 has been excluded in this figure due to the poor SED fitting at the center of this source as discussed before (and in Appendix~\ref{app:C25}). 

We compare our results with the scaling relations between $L_{\rm [CII]}$ and SFR found by \cite{DeLooze2014}, \cite{Herrera-Camus2015}, and \cite{Schaerer2020}. To facilitate the comparison, for each galaxy, we fit our resolved scaling relation using a power law ($\rm \Sigma_{SFR_{SED}} \propto {\Sigma_{L_{[CII]}}^\alpha} $; we exclude the outer regions of the galaxies at radii $>4$\,kpc to avoid possible contamination by extended emission, as discussed below). Overall, our data follow the relations of \cite{DeLooze2014} and \cite{Schaerer2020}, which are based on integrated measurements for intermediate- and high-redshift galaxies, respectively. Interestingly, our results tend to be more offset from the relation in \citet{Herrera-Camus2015}, which is based on resolved measurements at $z\sim0$, in the sense that our spatial bins tend to have higher SFR per unit [CII] than what would be predicted by that relation. A previous spatially-resolved study of kpc-scale star-forming clumps in $z>5$ galaxies by \cite{Carniani2018} also found a similar offset. They attribute this to a combination of lower gas metallicity (leading to both lower carbon abundance and less photoelectric dust heating of the gas), increased gas ionization parameters which suppresses [CII] emission, and extended [CII] emission in their galaxies compared (see also, e.g., \citealt{Ferrara2019,Markov2022}).

We find a median slope of the $\rm \Sigma_{SFR_{SED}} - {\Sigma_{L_{[CII]}}}$ relation of $\alpha=1.58\pm0.02$ for our sample. For about half of our sources, this relation is markedly steeper than the literature ones.
We note that, for sources such as CRISTAL-11 and CRISTAL-21, the slopes would be even steeper if we had included the outer regions in the fitting, due to excess [CII] emission compared to the SFR in those regions. 
Two factors may influence this slope: the existence of a [CII] deficit in the central regions, where the SFR density is higher (e.g., \citealt{Lutz2016,Herrera-Camus2018ApJ,Liang2024}), or spatially-extended emission that does not directly trace star-forming regions (e.g., \citealt{Fujimoto2019,Rybak2019}). Both could explain the steeper slopes found for some of our sources. A [CII] deficit in the central regions, where the SFR density is higher, could be caused by the intense radiation fields in those regions (e.g., \citealt{Smith2017,Rybak2019}) and/or lower gas metallicities and densities (e.g., \citealt{Smith2017,Ferrara2019}). Lower central gas metallicities have been found for dwarf galaxies at $z\simeq2$ by \cite{Wang2019}, which they attribute to feedback-driven outflows. This would agree with the scenario put forth in Section~\ref{sec:fobs-r} in which the central regions of our sources might be less dust-obscured because of outflows. Spatially-resolved measurements of the gas metallicities with JWST would be needed to investigate whether our sources also present signs of inverted metallicity gradients such as the ones found by \cite{Wang2019}. At the same time, most of our CRISTAL galaxies show extended [CII] emission compared to both the UV and IR emission (\citealt{Ikeda2024}; see also \citealt{Herrera-Camus2021,Solimano2024,Posses2024}), so this could also be contributing to our steeper slopes.

We conclude that, while empirical relations calibrated on whole galaxies might be useful to obtain the spatial distribution of SFR in galaxies from resolved [CII] observations, they may not hold for galaxies with [CII] halos or outflows, especially in the outskirts. A more detailed study of resolved SFR--$\rm L_{[CII]}$ relation in CRISTAL galaxies will be presented by Palla et al. (in prep.).

\section{Conclusion}\label{sec:conclusion}

In this work, we developed a spatially-resolved approach to study UV- and submm-detected main-sequence star-forming galaxies at $4<z<6$ from the CRISTAL sample. All sources in our sample have HST, JWST/NIRCam and ALMA Band 7 observations at resolutions equal to or better than 0.5\arcsec. We prepared these multi-wavelength data sets by astrometrically aligning and PSF-matching them. Then, we measured spatially-resolved multi-wavelength photometry on an average of 195 spatial bins per galaxy, each corresponding to a scale of $\sim0.5-1$\,kpc. Each spatially-resolved SED was then modeled independently using the \magphys\ code to obtain the spatial distributions of physical properties related to the star formation activity and dust content of our galaxies. For each CRISTAL galaxy, we obtain maps of SFR, stellar mass, specific SFR, average $V$-band dust attenuation, and total dust luminosity. We also computed the SFR and the fraction of dust-obscured star formation using empirical, SED-model-independent prescriptions for comparisons with previous studies and consistency checks. We summarize our main findings below.

\begin{itemize}

\item Our SED modeling approach is valid on sub-galactic scales at high-redshift. The resolved SEDs of all CRISTAL galaxies are well fitted, except for a small fraction ($5\%$) of poorly fitted SEDs in the low-surface brightness regions in galaxy outskirts. Despite each spatial bin being fitted independently from adjacent ones, our \magphys\ physical parameter maps are free of strong discontinuities and noise (the only exception is the central region of CRISTAL-25, which we speculate could be contaminated by an AGN; Appendix~\ref{app:C25}).

\item All our CRISTAL galaxies show a strong correlation between the surface density of stellar mass, \sigmstar, and the surface density of star formation rate, \sigsfr, on $\sim0.5-1$\,kpc scales, i.e., we recover a resolved star-forming main sequence (rSFMS) at $z\simeq5$. This rSFMS has a typical slope of $0.85\pm0.02$ and scatter of $\sim0.1$dex, similar to the observed rSFMS in local and cosmic noon galaxies at similar physical scales. Taken together with previous works at $z\simeq0$ (e.g., \citealt{pessa2021}) and $z\simeq1-2$ (e.g., \citealt{Wuyts2012,Wuyts2013}), our results show that a sub-galactic resolved main sequence persists from the local Universe through cosmic noon and all the way to $z\sim5$ in typical star-forming galaxies, despite significant changes in the stellar populations and dust obscuration properties with cosmic time.

\item We explore the well-known relation between the fraction of dust-obscured star formation, \fobs, and the stellar mass, on spatially-resolved scales. These two properties are positively correlated on global scales in galaxies from the local to the high-redshift Universe (e.g., \citealt{whitaker2017,Algera2023}). Somewhat surprisingly, we do not find a clear positive trend between \fobs\ and \sigmstar\ at the resolved scales probed in our CRISTAL sample. If anything, \fobs\ and \sigmstar\ seem to be anti-correlated, with less obscured star formation in the central regions of galaxies where \sigmstar\ is higher. We speculate that this is due to stellar feedback  clearing the dust in the central regions, as the surface density of SFR also peaks in those regions. However, a strong caveat of our result is that our inferred dust luminosity relies on a single dust continuum measurement from ALMA, and could be sensitive to spatial dust temperature variations. This  needs to be further investigated by obtaining matched-resolution dust temperature maps (using high-frequency ALMA observations), as well as looking for evidence of molecular gas outflows and/or inverted metallicity gradients in our galaxies.

\item We demonstrate that \fobs\ correlates well with the average $V$-band dust attenuation constrained with \magphys, \av, for each spatial bin, as expected, and provide a simple analytical expression that relates \fobs\ and \av.

\item We test the consistency of results when performing SED fitting in a spatially-resolved manner compared to the traditional global, integrated approach. Contrary to some previous studies, we find no offsets between the total stellar masses derived from global photometry and the total stellar masses derived from adding up the stellar masses in all the spatial bins in a galaxy. That is, there is no evidence for significant outshining by young stellar populations. The same applies to the star formation rates, indicating no significant biases due to large concentrations of dust. This could be because the resolution of our observations is not high enough to reproduce offsets found in other studies, or simply because our \magphys\ modeling combined with ALMA observations can reliably break the age-dust degeneracy on small scales. We plan to explore this further in future studies.

\item Our sample is one of the few samples of high-redshift ($z>4$) star forming main-sequence galaxies for which both JWST and ALMA observations are available at high resolution. We test the impact on our results of having ALMA observations, probing the dust continuum emission. Specifically, we find that not including ALMA changes the radial distribution of sSFR in half of our galaxies, with local differences in sSFR that can be as high as a factor of three. \textcolor{black}{We highlight that ALMA observations can help breaking the age-dust degeneracy by informing the \ldust when producing maps of the stellar mass, dust attenuation, and star formation rate of galaxies.}
% We find that ALMA observations can be critical in breaking the age-dust degeneracy when producing maps of the stellar mass, dust attenuation, and star formation rate of galaxies. 
This can have significant impacts on the inferred spatial distribution of specific star formation rates, and, therefore, on conclusions about the assembly mode of galaxies from spatially-resolved studies. 

\item We use our spatially-resolved SFR maps to test empirical SFR prescriptions at $\lesssim1$\,kpc resolutions in our high-redshift galaxies. We find good correlations between our SED-based SFRs and SFRs based on the UV+IR fluxes, albeit with some scatter. We also find that our resolved SED-based SFRs correlate well with the [CII] luminosities measured by ALMA at least in the inner $\sim4$\,kpc regions of the CRISTAL galaxies. However, we caution that [CII] may not be the best SFR tracer on resolved scales, due to the large scatter (especially at large radii) and variable slopes of the resolved SFR-[CII] relation. This is likely due to excess [CII] flux from extended emission \citep{Ikeda2024} as well as [CII] deficits in the central regions. Our resolved SFR-[CII] relation is offset from that at $z\simeq0$ found by \cite{Herrera-Camus2015}, confirming that high-redshift galaxies seem to emit less [CII] per unit SFR even on resolved scales. This is likely due to different local ISM conditions, such as lower metallicities and stronger radiation fields. Additional metallicity and gas heating diagnostics from JWST and ALMA should help elucidate this in the future.
\end{itemize}

This work demonstrates the potential of spatially-resolved, multi-wavelength analyses in galaxies beyond cosmic noon, enabled by the combination of HST, JWST, and ALMA, to start untangling the different physical components of galaxies. Such analyses were only possible in the local Universe and in dramatically-lensed high-redshift systems until recently. The physical parameter maps provided by this technique are highly complementary to other spatially-resolved studies. For example, they can be used to obtain the stellar mass distribution useful for kinematic studies, and the stellar mass and SFR distribution can be combined with maps of the cold gas distribution to investigate the resolved Kennicutt-Schmidt relation and the physical origin of the resolved star-forming main sequence.
\begin{acknowledgments}
We thank the referee for very thoughtful suggestions that helped us improve the submitted manuscript.
We thank Ian Smail, Simon Driver, and Aaron Robotham for useful comments and discussions.
EdC and JL acknowledge support from the Australian Research Council (projects DP240100589 and CE170100013).

AF acknowledges support from the ERC Advanced Grant INTERSTELLAR H2020/740120. JH acknowledges support from the ERC Consolidator Grant 101088676 (VOYAJ). I.D.L. acknowledges funding from the Belgian Science Policy Office (BELSPO) through the PRODEX project “JWST/MIRI Scence exploitation” (C4000142239), from the European Research Council (ERC) under the European Union’s Horizon 2020 research and innovation programme DustOrigin (ERC-2019-StG-851622) and from the Flemish Fund for Scientific Research (FWO-Vlaanderen) through the research project G0A1523N. RI is supported by Grants-in-Aid for Japan Society for the Promotion of Science (JSPS) Fellows (KAKENHI Grant Number 23KJ1006). MK was supported by the ANID BASAL project FB210003. MR acknowledges support from project PID2020-114414GB-100, financed by MCIN/AEI/10.13039/501100011033. MS was financially supported by Becas-ANID scolarship 21221511, and also acknowledges ANID BASAL project FB210003. KT acknowledges support from JSPS KAKENHI grant No. 23K03466. VV acknowledges support from the ALMA-ANID Postdoctoral Fellowship under the award ASTRO21-0062. 
Some of the data products presented in this work were retrieved from the Dawn JWST Archive (DJA). DJA is an initiative of the Cosmic Dawn Center, which is funded by the Danish National Research Foundation under grant No. 140.
We acknowledge that this paper was written on Noongar land, and pay our respects to its traditional custodians and elders, past and present.
This work used Astropy:\footnote{http://www.astropy.org} a community-developed core Python package and an ecosystem of tools and resources for astronomy \citep{astropy:2022}.
\textcolor{black}{The JWST and HST data presented in this article were obtained from the Mikulski Archive for Space Telescopes (MAST) at the Space Telescope Science Institute. The specific observations analyzed can be accessed via \dataset[10.17909/se5h-8m79]{\doi{10.17909/se5h-8m79}}}
\end{acknowledgments}

%\vspace{5mm}

\facilities{ESO/ALMA, STScI/HST, STScI/JWST}
\software{Astropy \citep{astropy:2013, astropy:2018, astropy:2022},  photutils \citep{photutils2022}, scipy \citep{scipy2020}, \magphys\ \citep{magphys2008,magphys2015,magphys2020}, WebbPSF \citep{Perrin2014SPIE}, TinyTim \citep{Krist2011SPIE}}

\bibliography{ref}{}
\bibliographystyle{aasjournal}
\appendix

\section{The intriguing case of source CRISTAL-25}\label{app:C25}

\begin{figure*}
    \includegraphics[width=\linewidth]{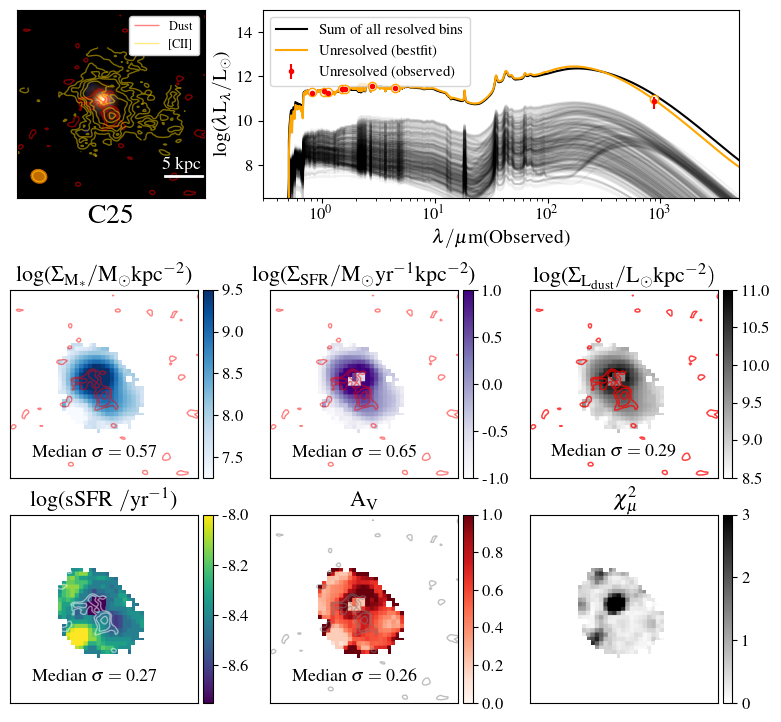}
    \caption{Resolved SED fitting results of CRISTAL-25 with all filters available included. Panels show the same properties as in Fig.~\ref{fig:maps}.}\label{fig:C25-maps}
\end{figure*}

\begin{figure*}
\center
    \includegraphics[width=0.65\linewidth]{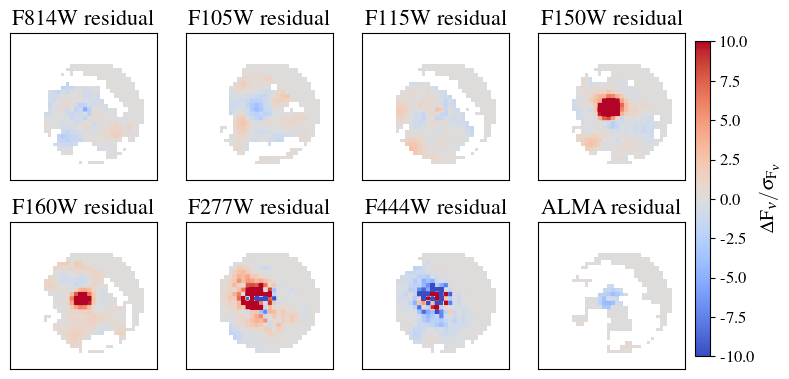}
    \includegraphics[width=0.3\linewidth]{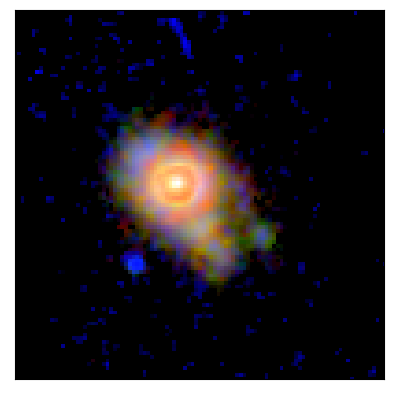}
    \caption{Left: CRISTAL-25 residual maps of all filters, shown as $\Delta F_\nu = (F_\nu^{\rm obs} - F_\nu^{\rm mod}) / \sigma_{\rm F_\nu}$, where $F_\nu^{\rm obs}$ is the observed flux in a given filter, $F_\nu^{\rm mod}$ is the best-fit model flux in that filter, and $\sigma_{\rm F_\nu}$ is the observational flux uncertainty. Right:RGB image of CRISTAL-25 made following the method in \citet{Lupton2004}. It is composed with F150W, F277W and F444W as B, G and R, respectively, before PSF-matching.}\label{fig:C25-residual}
\end{figure*}

% \begin{figure}
% \center
%     \includegraphics[width=0.3\linewidth]{C25-RGB.png}
%     \caption{RGB image of CRISTAL-25 made following the method in \citet{Lupton2004}. It is composed with F150W, F277W and F444W as B, G and R, respectively, before PSF-matching. }\label{fig:C25-RGB}
% \end{figure}

\begin{figure*}
    \includegraphics[width=\linewidth]{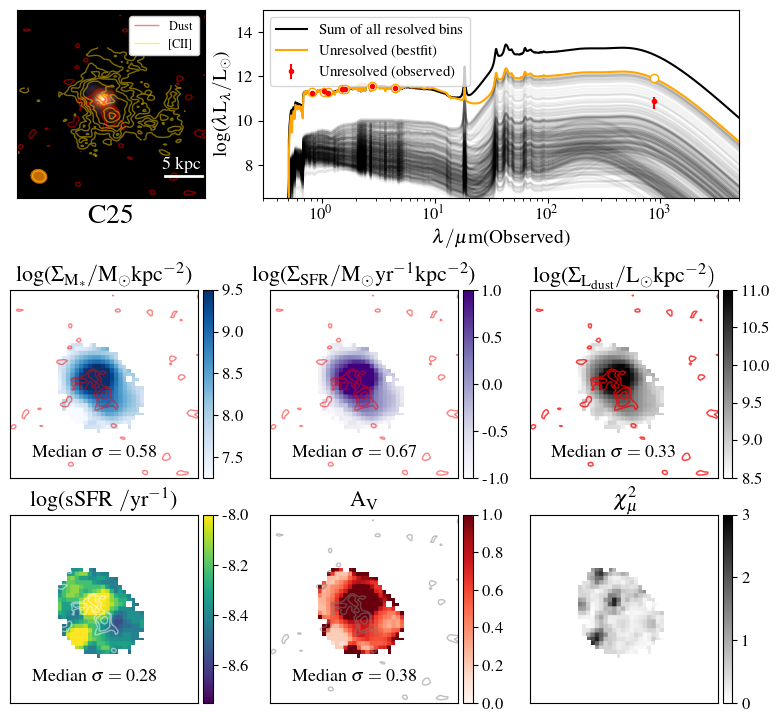}
    \caption{Resolved SED fitting results of CRISTAL-25 without ALMA observation {\em included}. Panels show the same properties as in Fig.~\ref{fig:maps}.}\label{fig:C25-maps-noALMA}
\end{figure*}

As discussed in the main text, our resolved SED fitting performs poorly in the central region of CRISTAL-25. This can be seen in Fig.~\ref{fig:C25-maps}, where the central region, for which we have the highest SNR, shows the highest $\chi^2_{\mu}$ values, indicating a poor fit. This is not seen in any other CRISTAL galaxies: for the rest of the sample, higher $\chi^2_{\mu}$ values tend to appear in the outskirts, where SNR are typically lower and the observations more uncertain.
For CRISTAL-25, the inferred SFR, dust luminosity, and dust attenuation maps also show strong spatial discontinuities, in which the central part appears to be quenched, with very low SFR ($\sim10^{-1}\,\rm M_\odot yr^{-1}$) and very low sSFR ($\sim10^{-10}\, \rm yr^{-1}$). We deem these results unreliable due to the $\chi^2_{\mu}$ behaviour.

To investigate this, we generate fit residual maps for all the filters used, shown in Fig.~\ref{fig:C25-residual}. These maps show markedly large residuals in the central region, especially in the filters from F150W to F444W. The residuals (observation minus model) are positive for F150W, F160W and F277W, but negative for F444W. We note that JWST/F150W and HST/F160W cover similar wavelengths but come from different telescopes, yet they show similar excess in fluxes in the central region. This suggests that the residuals come from true mismatches between the SED model and the observations, and not from systematics in the flux calibration.

We further investigate this by analysing the JWST image of this source (before PSF homogenization). A dim central point source can be clearly identified in the RGB image in Fig.~\ref{fig:C25-residual} which requires a steeper stretch to notice.
We suspect that the flux excess could be caused by broad emission lines from an active galactic nucleus (AGN), because the flux excess appears in filters which could be contaminated by [MgII] $\lambda2798$\,\AA\ (i.e., F150W and F160W) and H$\beta$+[OIII] (i.e., F277W) at the redshift of this source ($z=4.573$). 
This hypothesis will be tested with the upcoming JWST Cycle 2 GO program (ID:3045; PI: A. Faisst) which will observe 18 galaxies (including CRISTAL-25) with the NIRSpec IFU using F170LP and F290LP filters at $R\sim1000$. At the redshift of CRISTAL-25, this program can observe the wavelengths of H$\beta$+[OIII] and H$\alpha$+[NII] at different locations of the galaxy. If there are broad line features in the central region, it will can confirm the presence of a (Type 1) AGN in this source.
If CRISTAL-25 is found to host an AGN, this suggests that our method has the potential to detect and identify these objects while integrated studies failed to even spot any abnormalities. 
The relative contribution from the AGN is high enough to be noticeable only in the central resolution element of our analysis, causing poor SED fitting results. The total (integrated) SED does not show significant contamination.
Therefore, integrated studies could not identify the potential AGN that is outshined by its host galaxy.

Another potential reason why previous SED studies \citep{Mitsuhashi2023,weaver_cosmos2020_2022,laigle_cosmos2015_2016} of this galaxy did not identify a potential AGN was the lack of ALMA observations. To demonstrate this, in Fig.~\ref{fig:C25-maps-noALMA}, we present the results of our resolved SED fitting of CRISTAL-25 in a run where we excluded the ALMA observation. In this case, we find that the $\chi^2_{\mu}$ values improve in the central region. This is because the model now includes a dusty starburst (high \av, high sSFR) to explain the optical colors in the central region --  entirely the opposite to the case with ALMA observations. This central dusty starburst can be ruled out because it would severely overshoot the ALMA dust continuum flux, as shown in Fig.~\ref{fig:C25-maps-noALMA}, where sum of the spatial bin SEDs obtained from only fitting the HST+JWST data would produce an ALMA Band 7 dust continuum flux much higher (by two orders of magnitude) than the measurement (which would effectively put CRISTAL-25 in the sub-millimetre galaxy regime). Even a fit to the integrated HST+JWST photometry alone still overpredicts the ALMA flux by a factor of 10.

This example demonstrates again the importance of having FIR observations to break the age-dust degeneracy.
%With resolved approach, we can rule out the dusty star burst scenario with the (un-fitted) ALMA observation since a few central resolved SED each predicts a 870$\mu$m flux higher than the total flux of the whole galaxy. The summation of all bestfit resolved SED indicates an even higher SFR of $\sim10^3\rm M_{\odot}yr^{-1}$ which could be in the regime of sub-millimeter galaxies (SMGs).
%This example demonstrates again the importance of having FIR observation to break the dust-age degeneracy. 
One may have concluded that CRISTAL-25 is (nearly) as SMG with a central dusty starburst based on pixel-by-pixel SED fitting including only HST and/or JWST observations, as is now routinely done in the literature (e.g., \citealt{Abdurrouf2022, GimenezArteaga2023,Song2023}).
Even with low resolution, the integrated flux at FIR/sub-millimetre wavelengths can still be used to compare with the model predicted flux to verify the stellar properties inferred from UV/optical/NIR filters alone (e.g., \citealt{Smail2023}).

%If high-resolution FIR observations are available, CRISTAL-25 shows that spatially-resolved SED fitting might allow one to detect faint central AGN -- though this needs to be confirmed with more observations.

%This would lead to a higher integrated SFR ($\log({\rm{SFR}/$\rm{ M_{\odot}yr^{-1}})=2.89^{+0.23}_{-0.21}$), than when we include ALMA ($\log({\rm{SFR}/$\rm M_{\odot}yr^{-1})=2.57^{+0.11}_{-0.15}$).  
%The higher SFR in the ALMA-dropping case is found closer to the integrated SED fitting in \citet{Mitsuhashi2023} of CRISTAL-25 using CIGALE with COSMOS survey photometry \citep{laigle_cosmos2015_2016}, which yields log(SFR/$\rm M_{\odot}yr^{-1})=2.75\pm0.29$.\\

\section{Dependence of the dust obscuration on the dust temperature}\label{app:Tdust}

\begin{figure}
    \centering
    \includegraphics[width=0.85\linewidth]{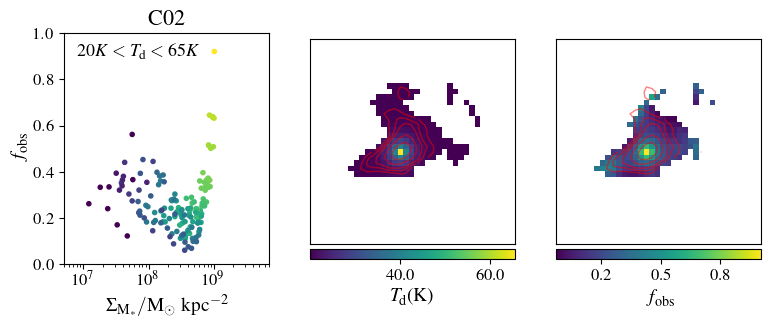}
    \caption{Left: \fobs\ vs \sigmstar\ for CRISTAL-02 assuming the 2D $T_{\rm dust}$ profile shown in the central panel. Center: 2D Sersic $T_{\rm dust}$ profile with $n=1$ and $R_e=0.25$\arcsec\ \citep{cochrane2019}. The dust temperature profile peaks where the dust emission peaks, and is shifted up to achieve a maximum temperature $\sim65$\,K and minimum temperature of $20$\,K. Right: \fobs\ map of CRISTAL-02 with the $T_{\rm dust}$ profile, generated using an optically-thin modified black-body dust emission model with emissivity index $\beta=2$. The red contours show the observed ALMA Band 7 dust continuum.}
    \label{fig:fobs-Tdust}
\end{figure}

Here we investigate how the inferred spatial distribution of dust-obscured star formation, \fobs, depends on spatial variations of the dust temperature. We use CRISTAL-02 as an example. We simulate a dust temperature gradient using an isothermal optically-thin modified black-body (MBB) dust emission model (e.g., \citealt{daCunha2021}) for each spatial bin. We adopt the dust temperature radial profile from \citet{cochrane2019}, which follows an exponential profile with effective radius 0.25\arcsec, and a dust temperature varying from 20 to 65\,K, peaking at the center of the dust emission (middle panel of Fig.~\ref{fig:fobs-Tdust}). Then, we normalize the MBB dust emission spectrum to the observed ALMA Band 7 flux in each spatial bin, and integrate it to obtain the dust luminosity, which converts to $\rm \Sigma_{SFR_{IR}}$. Finally, we combine with the UV SFR to estimate \fobs\ in each spatial bin, as described in Section~\ref{sec:fobs-trend}. The resulting \fobs\ map is shown in the third panel of Fig.~\ref{fig:fobs-Tdust}, and the resulting relation between \fobs\ and \sigmstar\ is shown in the first panel of Fig.~\ref{fig:fobs-Tdust}. We find that, that despite the increase of \fobs\ in the central region increased due to the higher central temperature, it only complicates the previous trend (Fig.~\ref{fig:fobs-Mstar}), as only the hotter central spatial bins deviate from the anti-correlation. Most of the outer regions only have mild temperature variation (i.e., $20-30$\,K), and the negative trend between \fobs\ and \sigmstar remains.

\end{document}